\documentclass{ieeeaccess}
\usepackage{cite}
\usepackage{amsmath,amssymb,amsfonts}
\usepackage{algorithmic}
\usepackage{graphicx}
\usepackage{textcomp}
\usepackage[hyperindex,pdftex]{hyperref}
\usepackage{float}
\usepackage{subcaption}
\usepackage{booktabs,caption,makecell}
\usepackage[flushleft]{threeparttable}
\usepackage{booktabs}
\usepackage{epsfig}

\usepackage{multirow}
\begin{document}
\doi{}

\begin{titlepage} 

	\centering 
	
	\scshape 
	
	\vspace*{\baselineskip} 
	
	
	\rule{\textwidth}{1.6pt}\vspace*{-\baselineskip}\vspace*{2pt} 
	\rule{\textwidth}{0.4pt} 
	
	\vspace{0.75\baselineskip} 
	
	{\LARGE Landscape of Big Medical Data:\\A Pragmatic Survey on Prioritized Tasks\\} 
	
	\vspace{0.75\baselineskip} 
	
	\rule{\textwidth}{0.4pt}\vspace*{-\baselineskip}\vspace{3.2pt} 
	\rule{\textwidth}{1.6pt} 
	
	\vspace{2\baselineskip} 
	
	
	
	\vspace*{2\baselineskip} 
	
	
	Edited By
	
	\vspace{0.5\baselineskip} 
	
	{\scshape\Large Zhifei Zhang\\Wanling Gao\\Fan Zhang\\Yunyou Huang\\Shaopeng Dai\\Fanda Fan\\Jianfeng Zhan\\Mengjia Du\\Silin Yin\\Longxin Xiong\\Juan Du\\Yumei Cheng\\Xiexuan Zhou\\Rui Ren\\Lei Wang\\Hainan Ye\\} 
	
	\vspace{0.2\baselineskip} 

	\vfill 
	
	
	\epsfig{file=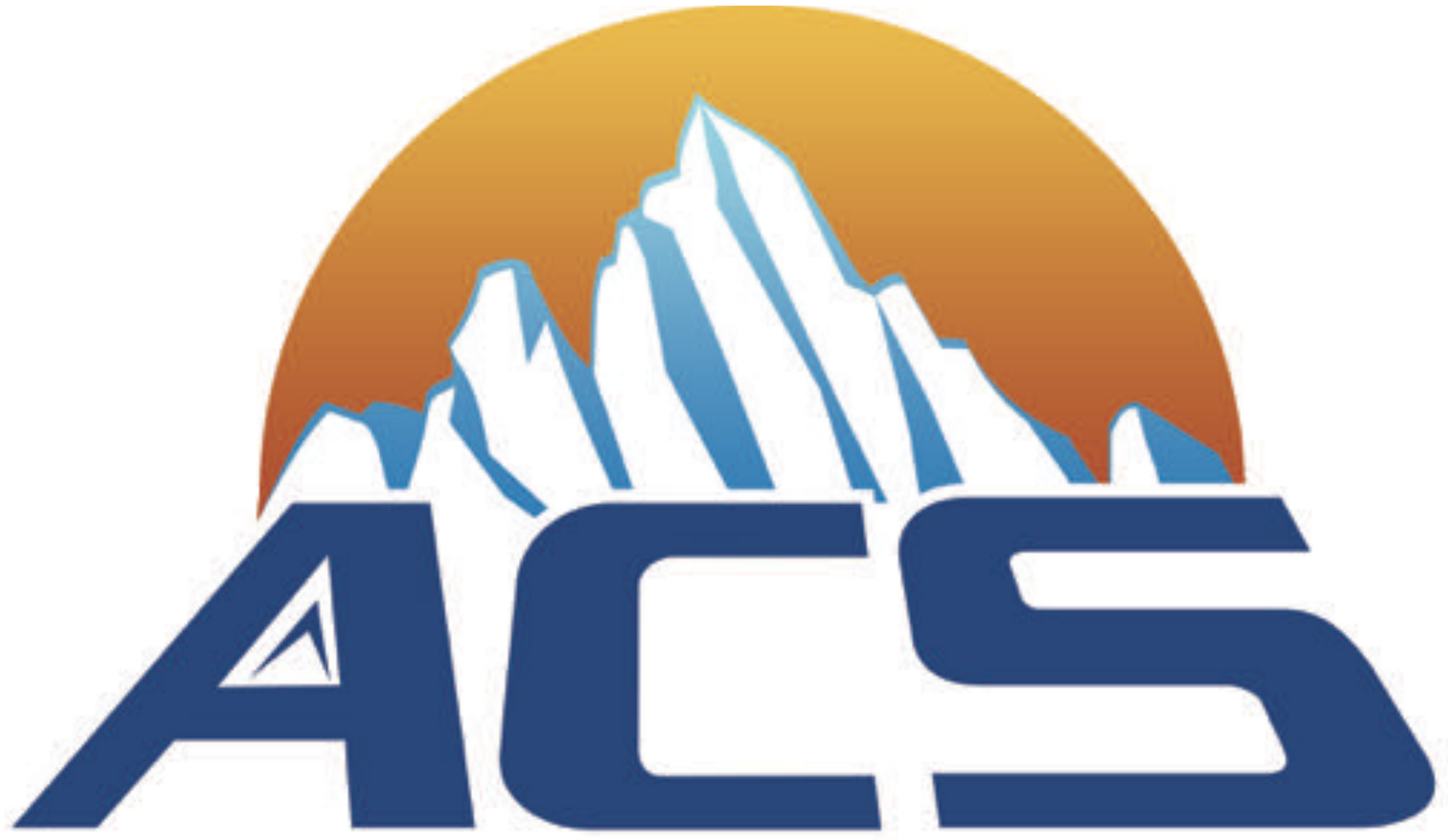,height=5cm,angle=270}
	\textit{\\Department of Physiology and Pathophysiology,
Capital Medical University\\Software Systems Laboratory (SSL), ACS\\ICT, Chinese Academy of Sciences\\Beijing, China} 
	\vspace{3\baselineskip} 

	Technical Report No. ACS/SSL-2019-1 
	
	{\large January 3, 2019} 

\end{titlepage}

\title{Landscape of Big Medical Data: A Pragmatic Survey on Prioritized Tasks}
\author{\uppercase{Zhifei Zhang}\authorrefmark{1},
\uppercase{Wanling Gao\authorrefmark{2,3}, Fan Zhang\authorrefmark{2}, Yunyou Huang\authorrefmark{2,3}, Shaopeng Dai\authorrefmark{2,3},  Fanda Fan\authorrefmark{2,3}, Jianfeng Zhan\authorrefmark{2,3}, Mengjia Du\authorrefmark{2,3}, Silin Yin\authorrefmark{2}, Longxin Xiong\authorrefmark{4}, Juan Du\authorrefmark{5}, Yumei Cheng\authorrefmark{6}, Xiexuan Zhou\authorrefmark{2,3}, Rui Ren\authorrefmark{2,3}, Lei Wang\authorrefmark{2},  Hainan Ye\authorrefmark{7}}.
}
\address[1]{Department of Physiology and Pathophysiology,
Capital Medical University}
\address[2]{State Key Laboratory of Computer Architecture, Institute of Computing Technology, Chinese Academy of Sciences}
\address[3] {University of Chinese Academy of Sciences}
\address[4]{Nanchang First Hospital}
\address[5]{Neonatal Center Beijing Children's Hospital, Affiliated to Capital Medical University}
\address[6]{Beijing Obstetrics and Gynecology Hospital, Affiliated to Capital Medical University}
\address[7]{Beijing Academy of Frontier Sciences and Technology}
\tfootnote{This work is supported by the National Key Research and Development Plan of China (Grant No. 2016YFB1000600, 2016YFB1000605, and 2016YFB1000601).}

\markboth
{Zhang \headeretal: Landscape of Big Medical Data: A Pragmatic Survey on Prioritized Tasks}
{Zhang \headeretal: Landscape of Big Medical Data: A Pragmatic Survey on Prioritized Tasks}

\corresp{Corresponding author: Zhifei Zhang (e-mail: zhifeiz@ccmu.edu.cn).}

\begin{abstract}
Big medical data poses great challenges to life scientists, clinicians,  computer scientists, and engineers.
In this paper, a group of life scientists,  clinicians, computer scientists and engineers sit together to discuss several fundamental issues. First, what are the unique characteristics of big medical data different from those of the other domains? Second, what are the prioritized tasks in clinician research and practices utilizing big medical data? And  do we have enough publicly available data sets for performing those tasks? Third, do the state-of-the-practice and state-of-the-art algorithms perform good jobs? Fourth, are there any benchmarks for measuring algorithms and systems for big medical data?   Fifth, what are the performance gaps of state-of-the-practice and state-of-the-art   systems handling big medical data currently or in future? Finally but not least, are we, life scientists,  clinicians, computer scientists and engineers,  ready for working together?
We believe answering the above issues will help define and shape the landscape of big medical data.
\end{abstract}


\titlepgskip=-15pt

\maketitle

\section{Introduction}
\label{sec:introduction}

Unlike physics or chemistry, which the natural laws governing molecules are successful in describing~\cite{tatonetti2017translational}, medical science is not founded on first principals from which a healthy or unhealthy human being or animal can be derived.  Thus in nature, one of the important features of medical science  is its data-driven mode: massive medical data stems from a wide range of experiments or clinical practices that spit out many types of information~\cite{marx2013biology}, and they provide the basis for our clinician research and practice.

Big medical data  poses great challenges to life scientists, clinicians,  computer scientists, and engineers. Even only considering computing requirements without delving into medical details, Stephens et al.~\cite{stephens2015big}  compared genomics data---one portion of big medical data, with three other major data sources: astronomy, YouTube, and Twitter, and concluded big medical data is either on par with or the most demanding of the domains in terms of data acquisition, storage, distribution, and analysis~\cite{stephens2015big}. Unfortunately, big medical data has many other dimensions of complexity other than data volume. For example, medical data is much more heterogeneous than those in the other domains~\cite{marx2013biology}. Taking  Alzheimer' s disease (AD)---the most common age-related neurodegenerative disease---as an example, clinicians and  researchers~\cite{adni-web} need collect
several types of data: clinical, genetic, imaging, and biospecimen data  for AD diagnosis. The heterogeneity of multi-source data not only raises cognition difficulty (for both clinician and computer scientist practitioners) , but also poses the challenges of managing and analyzing those data (for  computer scientists and engineers). The worst of all, the knowledge and skills in both fields are very professional, which seriously challenges multi-disciplinary collaboration.

\begin{figure}[!t]
\centering
\includegraphics[scale=0.12]{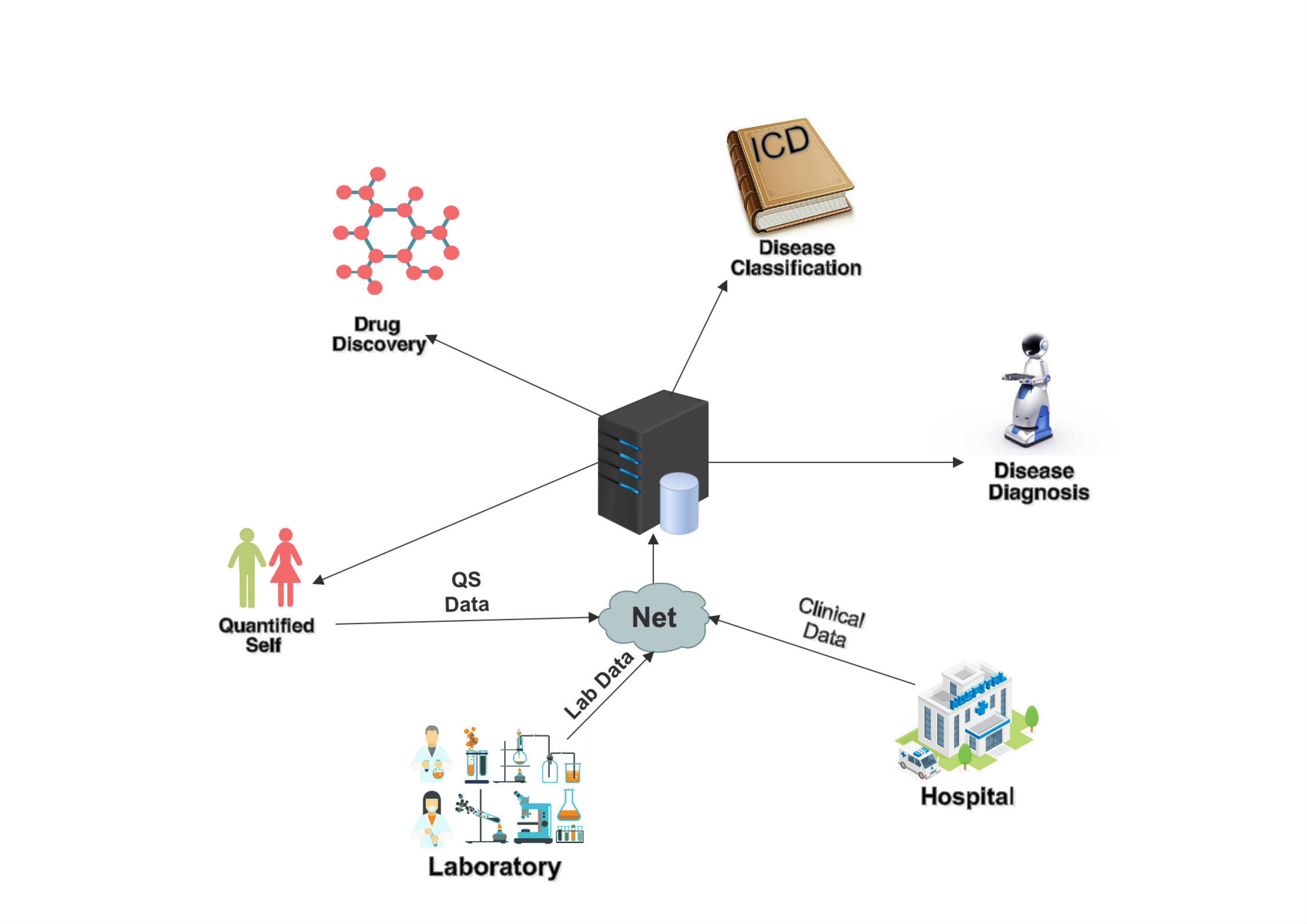}
\caption{Relationships of Prioritized Tasks with Different Medical Data.} 
\label{relationship}
\end{figure}

The purpose of this survey is to bridge the gap  among life scientists,  clinicians, computer scientists, and engineers. To define and shape the landscape of big medical data, we---a group of life scientists,  clinicians, computer scientists, and engineers---sit together. We know it is impossible to perform an exhaustive survey on all fields because of our knowledge and time budget limits. Instead, we take a pragmatic approach and focus on the prioritized tasks in clinician researches and practices: \emph{Quantified Self}---a specific movement to collect and analyze different aspects of a personal daily life; \emph{Disease Classification}; \emph{Disease Diagnosis}; and \emph{Drug Discovery}. Figure~\ref{relationship} summarize the relationships among different medical data with those prioritized tasks.

Also, our pragmatic approach lies in drafting this survey, and we keep the readers in our mind: for each prioritized task, we will help the readers---both  life scientists,  clinicians, computer scientists and engineers answer the following  questions. What data sets are publicly available? What are the state-of-the-practice and state-of-the-art algorithms and systems? Do they perform a good job? If not, how about the performance gap? Are there any comprehensive benchmark suite to evaluate the algorithms and systems?

As big medical data is a fast-evolving field, the another purpose of this survey is acting as a framework of defining and shaping the landscape of big medical data. For example, understanding the root causes of disease is an important task in utilizing medical data. Currently, we do not include them because of its complexity and immaturity.
Meanwhile, for one prioritized task---disease diagnosis, we only include three representative diseases to demonstrate how to utilize medical data: Alzheimer's disease (AD)---an age-related neurodegenerative disease, Acute lymphoblastic leukemia (ALL)---the most common cancer in children,  and breast cancer---the most common diseases among women. 
In one word, we will keep expanding  and updating state-of-the-art and state-of-the-practice of the full spectrum of big medical data.


Table~\ref{tab:medical_data_survey} performs a comprehensive comparison of our survey  with the previous ones. To the best of our knowledge, existing big medical surveys~\cite{shishvan2018machine,leveque2018state,razzak2018deep,ker2018deep,
liu2018benchmark,litjens2017survey,fang2016computational,kashyap2016big,
sharma2016medical,gani2016survey,islam2015internet} target  specific problems and thus fail to cover the whole spectrum of the above issues.
 Significantly different from the previous ones, our survey paper covers the four prioritized tasks in clinician research and practice: quantified self, disease classification, disease diagnosis, and drug discovery from perspectives of data sets, algorithms, systems, and benchmarks.

\begin{table*}[h!]
  \begin{center}
    \caption{Summary of medical data surveys}
    \label{tab:medical_data_survey}
    \begin{tabular}{c|c|c|c|c|c|c}
     \toprule
      \textbf{Reference} & \textbf{Publication year} & \textbf{Prioritized tasks}  & \textbf{Data sets}  & \textbf{Algorithms} & \textbf{Systems}&\textbf{Benchmarks}  \\
      \midrule
      Our & 2018 & Quantified self & \checkmark &  \checkmark & \checkmark &  \checkmark  \\
      {} & {} & Disease classification  &  \checkmark & \checkmark & \checkmark & \checkmark \\
      {} & {} & Disease diagnosis &  \checkmark &  \checkmark &  \checkmark &  \checkmark\\
      {} & {} & Drug discovery  &  \checkmark &  \checkmark & \checkmark & \checkmark\\

      \cite{shishvan2018machine} & 2018 & Quantified self &  \checkmark &  \checkmark &  \checkmark  & - \\
       {} & {} & Disease diagnosis & - &  \checkmark  &  \checkmark & - \\
      {} & {} & Drug discovery & - &  \checkmark & -  & - \\

      \cite{leveque2018state} & 2018 & Disease diagnosis  & \checkmark   & \checkmark & \checkmark & -  \\
      {} & {} & Disease surgery  & \checkmark   & \checkmark  & \checkmark & - \\

      \cite{razzak2018deep} & 2018 & Disease diagnosis &  \checkmark &  \checkmark &  \checkmark&  \checkmark \\

      \cite{ker2018deep} & 2018 & Disease diagnosis & \checkmark &  \checkmark &  \checkmark \\

      \cite{liu2018benchmark} & 2018 & Quantified self  & \checkmark  & \checkmark & \checkmark & \checkmark  \\

      \cite{litjens2017survey} & 2017 & Disease diagnosis  & \checkmark  & \checkmark & \checkmark & -  \\

      \cite{fang2016computational} & 2016 & Quantified self  & \checkmark  & \checkmark & \checkmark & -  \\
      {} & {} & Disease diagnosis  & \checkmark  & \checkmark & \checkmark & \checkmark  \\
      {} & {} & Drug discovery  & \checkmark  & \checkmark & - & -  \\

      \cite{kashyap2016big}  & 2016 & Quantified self  & - & \checkmark & \checkmark & -  \\
      {} & {} & Disease diagnosis  & \checkmark & \checkmark & \checkmark & -  \\

      \cite{sharma2016medical} & 2016 & Disease diagnosis  & \checkmark & \checkmark & \checkmark & \checkmark  \\

      \cite{gani2016survey} & 2016 & Big data index  & \checkmark  & \checkmark & \checkmark & \checkmark  \\

     \cite{islam2015internet} & 2015 & Quantified self   & -  & - & \checkmark & -  \\
     {} & {} & Drug discovery  & -  & - & - & -  \\

      \bottomrule
    \end{tabular}
  \end{center}
  \begin{tablenotes}
      \small
      \item  
    \end{tablenotes}
\end{table*}

After thoughtful discussion and comprehensive survey within our multi-disciplinary group, we gain several consensuses and insights as follows:
\begin{enumerate}
  \item Big medical data is heterogeneous, high-dimensional, embodying a large mixture of signals and errors~\cite{alyass2015big}. The widely used data mining or machine learning techniques heavily depend upon identifying weak  associations instead of strong causation. The noisy nature of experimental data may amplify the side effect of our current ability to identify weak associations at the cost of tolerating larger
error thresholds~\cite{alyass2015big}. From this perspective, it may indicate that we need develop new computing models and approaches in handling noisy big medical data.
  \item The publicly available big medical data sets  are limited in terms of  not only  its scale, but also its single data source. For example, massive previous work utilizes deep learning algorithms to analyze imaging data to automatically diagnose disease. Unfortunately,  all clinician practices and researches, like disease classification, disease diagnosis, and drug discovery need utilize comprehensive data sources. So, we need to build multi data-source  knowledge base  to advance state-of-the-art and state-of-the-practice for disease classification, disease diagnosis, and drug discovery.
  \item The previous work demonstrates the potential of incorporating machine learning techniques into clinician practices. However, its high accuracy is achieved on the static data. In reality, the clinician practitioners work in an open environment and handle open problems, so we need set up the realistic benchmarks that can mimic the way that the clinician practitioners handle the dynamic data for different clinician purposes. Or else, the achieved accuracy on static data does not make sense in the clinician practices.
  \item The sources and types of medical data are usually
multifarious and integrated. These dimensions are
not processed and learned individually, and conversely, they
are combined to detect and diagnose diseases cooperatively.
Under this circumstance, the storage and processing systems
are required to integrate different data sources and types.
 To the best of our knowledge, there exists no such a
system that can support multi-source and heterogeneous
data storage and processing in the big medical domain or even the other domains.
\item  We discuss several prerequisites for the purpose of an effective and efficient multi-disciplinary cooperation.
\end{enumerate}

In the following sections, the related terminologies and unique characteristics of big medical data are described in Section ~\ref{sec:terminologies} and Section~\ref{sec:unique_characterisitcs_medical_data}, respectively. In Section ~\ref{sec:prioritized_jobs}, we will discuss four prioritized jobs in clinician research and practises, and the related publicly available data sets.  In Section~\ref{sec:algorithm}, the state-of-the-art or state-of-practise algorithms are discussed. The  benchmark for measuring algorithms or systems will be shown in Section \ref{sec:bencmarks}. The performance gap of the state-of-the-art or state-of-practise systems handling big medical data will be explained in Section \ref{sec:gaps}. Also, we discuss multidisciplinary collaboration in Section~\ref{sec_collaboration}, and conclusions will be drawn in Section~\ref{sec_conclusion}.

\section{Terminologies}
\label{sec:terminologies}

For the readers with different background, this section explains several important terminologies in both medical sciences and computer sciences.

Genetics. Genetics~\cite{genome_gov}  is a term that refers to the study of genes and their roles in inheritance. As genes (units of heredity) carry the instructions for making proteins, directing the activities of cells and functions of the body, genetics involves scientific studies of genes and their effects~\cite{genome_gov}.

Omics. Omics~\cite{omics_wiki} aims at the collective characterization and quantification of pools of biological molecules that translate into the structure, function, and dynamics of an organism or organisms~\footnote{The English-language neologism omics~\cite{omics_wiki} informally refers to a field of study in biology ending in omics, such as genomics, proteomics or metabolomics.}.





High throughput technologies. The generic term refers to the technologies that allow exact and simultaneous examinations of thousands of genes, proteins and metabolites~\cite{blankenburg2009high}. For example, sequencing technologies~\cite{metzker2010sequencing} include a number of methods
that are grouped broadly as template preparation,
sequencing and imaging, and data analysis. The unique
combination of specific protocols distinguishes one
technology from another and determines the type of
data produced from each platform~\cite{metzker2010sequencing}.  In general, the automated Sanger method is considered as a first-generation technology~\cite{metzker2010sequencing}.  Second-generation sequencing (SGS) refers to sequencing of an
ensemble of DNA molecules with wash-and-scan techniques~\cite{schadt2010window}. Third-generation sequencing (TGS) refers to sequencing single
DNA molecules without the need to halt between read steps~\cite{schadt2010window}.


Systems biology.   The basic purpose of
systems biology~\cite{altaf2014systems} is the system-level understanding of a cell
or an organism in the context of
molecular networks. The four purposes of systems biology are as follows~\cite{altaf2014systems}: (1) understand the structure
of all the components of a cell/organism up to a molecular level.
(2) predict the future state of the cell/organism
under a normal environment, (3) predict the
output responses for a given input stimulus, and (4) estimate the changes in system behavior upon perturbation
of the components or the environment.

Comparative Medicine.
Comparative medicine~\cite{comparative_medicine_wiki} is a distinct discipline of experimental medicine that uses animal models of human and animal disease in translational and biomedical research. Basically, it relates and leverages biological similarities and differences among species to better understand the mechanism of human and animal disease~\cite{macy2017focus}.


Precision Medicine.   Precision medicine~\cite{mirnezami2012preparing} uses clinicopathological indexes and molecular profiling to create
diagnostic, prognostic, and therapeutic strategies individually tailored to a patient.

Personalized medicine. Personalized medicine~\cite{jain2002personalized} refers to the prescription of specific therapeutics that is the best suitable for
an individual on the basis of pharmacogenetic and pharmacogenomic information.

Data integration. The term data integration~\cite{gomez2014data}  refers to the situation where,
for a given system, multiple data sources are available and  studied integratively
for  knowledge discovery.

Benchmarks. A benchmark~\cite{fleming1986not} is the act of running  a set of computer programs or other operations, so as to assess the relative performance of an object, i.e., a system or an algorithm.

\section{What are the characteristics of Big Medical Data?}
~\label{sec:unique_characterisitcs_medical_data}


There are four significant characteristics of big medical data. 

First, big medical data is either on par with or the most demanding of the domains~\footnote{Stephens et al.~\cite{stephens2015big}  compared genomics data---one portion of big medical data, with three other major generators of Big Data: astronomy, YouTube, and Twitter.} in
terms of data acquisition, storage, distribution, and analysis~\cite{stephens2015big}. There are several reasons underlying the boom of big medical data.
On one hand, a significant
fraction of the world' s human population will have their genomes sequenced because of  the promise of precision medicine or personalized medicine~\cite{stephens2015big}.
On the other hand, for medicine, just having the
genome will not be sufficient, and other relevant
omics data sets will be definitely collected, i.e, transcriptome, epigenome, proteome, metabolome, and microbiome sequencing, from different tissues to compare healthy and
diseased states~\cite{chen2012personal}~\cite{soon2013high}~\cite{stephens2015big}.




Second, medical data is much more heterogeneous than those in the other domains~\cite{marx2013biology}. On one hand, they stem from a wide range of experiments that spit out many types of information~\cite{marx2013biology}.  In summary, extensively-used medical data consist of four types: sequence data, 3D-structure
data, multivariate data, and network data ~\cite{altaf2014systems}.
The challenge for integrating heterogeneous data lies in
deriving meaningful interpretable correlations and causation~\cite{alyass2015big}. For example,
direct correlation analyses between transcriptomics
and proteomics profiles are not valid in eukaryotic
organisms~\cite{alyass2015big}.
On the other hand, there are  diversity of existing data types and
formats, each one compliant to a different standard,
which results in data heterogeneity~\cite{gomez2014data}~\cite{goble2008state}.

Third, medical data is high-dimensional~\cite{alyass2015big}.
It is widely recognized
that multiple dimensions must be considered
simultaneously so as to  understand biological systems from a perspective of systems biology, and performing  analytics on high-dimensional data often results with poor interpretability~\cite{loos2012believe}. The reliability of models probably decreases with each
added dimension (i.e. increased model complexity) for a
fixed sample size (i.e. bias-variance dilemma)~\cite{clarke2008properties}. All estimate instability, model overfitting, local
convergence, and large standard errors compromise the
prediction advantage provided by multiple measures~\cite{alyass2015big}.

Fourth, medical data is coupled with the noisy nature of experimental data, and especially omics data embody a large mixture of signals and errors~\cite{alyass2015big}. The widely-used data mining or machine learning techniques heavily depend upon identifying weak associations instead of strong causation. The noisy nature of experimental data, which is  often found in Comparative Medicine, may amplify the side effect of our current ability to identify weak associations at the cost of tolerating larger
error thresholds~\cite{alyass2015big}. In other word, the popular data analytics techniques may fail in handling mixture of signals and errors.
Biological systems include non-linear
interactions and joint effects of multiple factors that
make it difficult to distinguish  signals from random
errors~\cite{alyass2015big}, which has two implications: first, it is  important to minimize sources of error with omics
data~\cite{alyass2015big}. Second, we need to develop new computing models and approaches in handling noisy big medical data.

\section{What are the prioritized tasks in clinician research and practices?}
~\label{sec:prioritized_jobs}

In this section, we summarize the prioritized tasks in clinician research and practices utilizing big medical data sets. We mainly focus on  four prioritized tasks: \emph{quantified self}, \emph{disease classification}, \emph{disease diagnosis}, and \emph{drug discovery}. For each task, we answer the same question: do we have enough publicly available data sets?

\subsection{Quantified Self}

So far, there is no uniform definition for quantified self (in short, QS). Wolf et al.~\cite{wolfqs} first propose QS in 2007, which uses technologies and equipments to record and analyze the body~\cite{almalki2015use, meleady2017surveillance, wolfram2012personal}. In this paper, we adopt the definition of QS in Wikipedia~\cite{wikiqs}---QS is a specific movement that integrates technology into data acquisition in a personal daily life~\cite{wikiqs}. We collect data in terms of biological, physical, behavioural, or environmental information, like food consumed, the quality of air, mood, and performance, whether mental or physical~\cite{wikiqs, zhu2018qs}. After analysing the data, QS equipments or apps will provide suggestions or warnings for users.


Zhu et. al~\cite{zhu2018qs} summarize four applications in QS, such as \emph{Daily Health Management}~\cite{van2015sleep,coughlin2016use,liu2015smartcare}, \emph{Disease Prevention}~\cite{barrett2013big,oresko2010wearable,axisa2005flexible}, \emph{Chronic Disease Management}~\cite{allet2010wearable,celler2003using,sun2015researchonwearable}, and \emph{Out-of-hospital rehabilitation}~\cite{bernard2002treatment,shou2015analysisofapp,fey1997out}. Personal data are continuously tracked and recorded through smart wearable devices~\cite{van2015sleep}. In daily life, tracking records are analyzed to provide suggestions and warning, which can also be shared among friends and families~\cite{coughlin2016use,liu2015smartcare}. Data collected by QS tools helps pinpoint the root cause of disease early and achieve early treatment to prevent disease~\cite{barrett2013big}. QS apps transfer the data to the hospital's medical record system and monitoring center, and provide early warning and corresponding diagnosis and treatment opinions in chronic disease~\cite{sun2015researchonwearable}. And with portable quantitative instruments, patients can monitor self at home and use computer software to ask doctors for advice remotely for regular review after surgery~\cite{fey1997out}.

\subsubsection{Publicly available data sets}

With the improvement of economic and the development of technology, QS  data is becoming more and more important in the field of health management and personalized medicine~\cite{swan2013quantified, jain2002personalized}.
However, most of QS data sets are limited to human behaviour. Moreover, the size of data set is small, and  only collected from several people or a dozen people. There is no comprehensively public access database for user-contributed data of QS~\cite{swan2013quantified}. The communities should establish a public repository where individuals could upload any types of QS data~\cite{hogg2013value}.
Several public database of QS data are as follows.

HAR (Human Activity Recognition) dataset is an important data category of QS, which is of great significance in improving walking stability, recognizing motor disorders, evaluating surgical outcomes, and reducing joint loading~\cite{shull2014quantified}.
At present, several HAR datasets have been released.
Daphnet freezing of gait dataset~\cite{bachlin2010wearable} is collected from patients with the Parkinson's disease. The dataset includes 10 users, including 1,140,835 samples~\cite{bachlin2010wearable}. The samples are marked as "frozen" or "non-frozen"~\cite{bachlin2010wearable}. WISDM Actitracker dataset~\cite{kwapisz2011activity} contains 1,098,213 samples belong to 28 users and 6 distinctive activities of sitting, jogging, walking, climbing, and standing stairs. They collected the data samples from Android phones~\cite{kwapisz2011activity}.  
Actitracker~\cite{lockhart2011design} contains several daily activities, including "jogging", "walking", "ascending stairs", and "descending stairs",and etc., which are collected using a cell phone in their pocket with 20 Hz sampling rate. Hand Gesture dataset~\cite{bulling2014tutorial} contains data on different types of  human hand movements. Table~\ref{tab211} shows seven public QS datasets.

The National Health and Nutrition Examination Survey (NHANES)~\cite{nhanes}, which can access the health and nutritional status, is a program designed for the United States (www.cdc.gov/nchs/nhanes/). The NHANES data set is available on Internet  through an extensive series of publications and articles in scientific and technical journals~\cite{nhanes}. Open Humans~\cite{openhumans} is a platform that allows you to upload, connect, and privately store your personal data -- such as genetic, activity, or social media data (www.openhumans.org/).

\begin{table}[htbp]
\caption{Overview of QS datasets. HAR means Human Activity Recognition. }
\begin{center}
\scalebox{0.78}[0.78]{
\renewcommand{\arraystretch}{1.5}
\begin{tabular}{|p{0.7in}|p{0.70in}|p{0.70in}|p{0.7in}|p{0.50in}|}
\hline
\textbf{Name} & \textbf{Task} & \textbf{Instance} & \textbf{Types} & \textbf{Year}   \\ \hline
Open Humans~\cite{openhumans} & Data Analysis & - & genetic, activity, or social media data & 2015-present \\ \hline
Hand Gesture~\cite{bulling2014tutorial} & HAR & sampling rate is 32 samples per second & time series data & 2014-present \\ \hline
WISDM~\cite{kwapisz2011activity} & HAR & 1, 098, 213 samples & time series data & 2011-present \\ \hline
Actitracker~\cite{lockhart2011design} & HAR & 29, 000 frames & time series data & 2011-present \\ \hline
Daphnet~\cite{bachlin2010wearable} & HAR & 1, 140, 835 samples & time series data & 2010-present \\ \hline
NHANES~\cite{nhanes} & Food Consumption & - & text & 1960-present \\ \hline
\end{tabular}
}
\label{tab211}
\end{center}
\end{table}

\subsection{Disease Classification}

Disease classification is a task that groups entities together according to their similarities, which is an important step in precision medicine and of great significance to the quantitative study of the medicine phenomenon~\cite{world1957manual, national2011toward}. Nowadays, it has been widely used in academic medicine~\cite{wang2017classification}. However, existing disease taxonomy is often focused on physiological characterizations and clinical appearance of diseases, with little reference to the diseases mechanism~\cite{bell1998new,park2017towards}. Constructing a more accurate classification of the diseases that can reflect the diseases mechanism is essential towards fully understanding the entities.

With the development of the biotechnology, computer technology and medical technology, the molecular data is growing rapidly. This data provides new knowledge of the diseases on a molecular level, and deepens our understanding of the disease. Thus, we can reclassify the diseases according to the new input---the molecular biology information (such as the genomic data of the tumour).



\subsubsection{Publicly available data sets}

Encouraged by the concept of precision medicine, most of the researches on reclassifying the diseases focus on the molecular biology. Besides, according to~\cite{NEJMp1114866}, for the purpose of building a new effective system of disease taxonomy, other comprehensive data concerning clinical medicine, environment and the state of individual health, which is expected to be accumulated, is also indispensable.  As shown in Table~\ref{tab:table1}, there are many publicly available data sets which are qualified for disease classification. So far the publicly available data mainly includes the genomic information of diseases like different kinds of cancers and clinical reports. However, the datasets relating to the environment and the state of individual health are rarely publicly accessible since they involve the patient privacy. Several representative public datasets containing genome, clinical medicine, circumstance and the health status of patients are described as below.

\newcommand{\tabincell}[2]{\begin{tabular}{@{}#1@{}}#2\end{tabular}}  
\begin{table*}[h!]
  \begin{center}
    \caption{Data set for disease classification}
    \label{tab:table1}
    \begin{tabular}{|p{0.7in}|p{2in}|p{1.3in}|p{1.1in}|p{0.70in}|}
     \toprule
      \textbf{Name} & \textbf{Task} & \textbf{Instances} & \textbf{Types}  & \textbf{Year} \\
       \midrule
      APGI~\cite{APGI2018}&\tabincell{c}{Expediting the transformation from research\\ detection to the improvement of the \\treatment  for pancreatic cancer patients.}&\tabincell{c}{More than  4,000 cases}&\tabincell{c}{genomic and clinical data}&\tabincell{c}{2009 - present}\\
      \midrule
      EGA~\cite{lappalainen2015european}&\tabincell{c}{Offering free data and services of biological\\ information.}&\tabincell{c}{Over 6.25 million cases}&\tabincell{c}{Data of phenotype and\\ gene}&\tabincell{c}{2007 - present}\\
      \midrule
      ICGC~\cite{ICGC2018}&\tabincell{c}{Describing cancer genomes.}&1.52 PB&\tabincell{c}{genomic and clinical data}&\tabincell{c}{2007 - present}\\
      \midrule
     TCGA~\cite{TCGA2018}&\tabincell{c}{Enhancing the avoidance, diagnosis, and \\therapy of cancer.}
&33096 cases&genomic data&\tabincell{c}{2000 - present}\\
       \midrule
      GEO~\cite{GEO2018}&\tabincell{c}{Offering a robust, multi-functional database\\ and effective tools to the query and \\analysis of  the database.}&\tabincell{c}{2,680,676  samples}&genomic data&\tabincell{c}{2000 - present}\\
      \midrule
      CCLE~\cite{CCLE2018}&\tabincell{c}{A transformation  from genomics of cancer\\ cell into cancer division.}&\tabincell{c}{Over 1100 cell lines contain-\\ing genetic information.}&\tabincell{c}{genomic and cell lines \\data}&\tabincell{c}{2000 - present}\\
      \midrule
      Sanger\cite{Sanger2018}&\tabincell{c}{Genomic detection and comprehension\\ on the Earth.}&\tabincell{c}{More than  1000 human \\ Genomes}&\tabincell{c}{genomic data }&\tabincell{c}{1992 - present}\\
      \midrule
      \tabincell{c}{Truven \\Marketscan~\cite{IBM2018}}&\tabincell{c}{Providing health improvement resolutions set \\up from complete data, progressive \\analysis and related specialists.}
&\tabincell{c}{Roughly 55 million claim \\every year.}&\tabincell{c}{genetic, environmental \\and other medical \\insurance data}&\tabincell{c}{1970s - present}\\

      \bottomrule
    \end{tabular}
  \end{center}
\end{table*}

\textbf{The Cancer Genome Atlas data set}

The Cancer Genome Atlas (TCGA) data set is the most popular genetic dataset, which focuses on collecting the data of the cancer patients. So far, TCGA has collected approximately 7000 human tumors~\cite{TCGA2018}. And, it contains many types of data such as measurements of somatic mutations, copy number variation, mRNA expression, et al~\cite{TCGA2018}. One of the most important purpose of TCGA is to obtain the valuable insights of the heterogeneity of different cancer subtypes~\cite{akbani2014pan,weinstein2013cancer,cancer2008comprehensive,cancer2012comprehensive}.


\textbf{European Genome-phenome Archive(EGA)}

EGA provides over 4 thousand datasets, consisting of individually distinguishable data in phenotype and gene. Those data are collected from the research of biomedicine, and can merely be used for legally genuine research purpose with permission by creating accounts~\cite{lappalainen2015european}. Specifically, EGA contains approximately 710 DNA samples without plasma cell and 428 white blood cell samples collected from over 4 hundred patients with metastatic prostate cancer.

\textbf{Australian Pancreatic Cancer Genome Initiative (APGI) data set}

 APGI is a part of the International Cancer Genome Consortium---a global research enterprise of over 100 scientists, clinicians and allied health professionals involved in pancreatic cancer research and care~\cite{APGI2018}.
APGI contains over 4,000 pancreatic cancer patients in the database, with a range of prospectively collected and archived biospecimens~\cite{APGI2018}. Every biospecimen is coupled with detailed clinico-pathological data, including past medical history, treatment data and detailed disease outcomes~\cite{APGI2018}.

\textbf{Cancer Cell Line Encyclopedia (CCLE) data set}

CCLE is a cooperation among Novartis Institutes for Biomedical Research, Genomics Institute of the Novartis Research Foundation, and the Broad Institute\cite{CCLE2018}. The project has three purposes: portray the features of gene and pharmacology of a large amount of cancer, promote synthetic analysis of connecting special susceptibility of pharmacology concerning the patterns of genome, and transform comprehensive genomics of cancer cell into division of human cancer~\cite{CCLE2018}. The web site~\cite{CCLE2018} provides publicly available genetic dataset that can be displayed in visualized methods. The data set,  involving more than 1 thousand related information of cell lines, can be exploited to quantitative analysis of cancer and research of cancer reclassification according to gene and cell lines.

\textbf{Truven Marketscan data sets}

Truven Marketscan is a synthetic platform of the IBM Watson Health$^{TM}$ business providing health improvement resolutions~\cite{IBM2018}. The clinical information of patients and the accumulation of over 50 million medical insurance claims every year are included, some of which, according to the description of the platform, are accessible for the aim of relative research with a fee. The medical insurance information can be applied to various aspects of medical research, especially disease classification concerning precise medicine due to the genetic and environmental information of plenty of patients.

\subsection{Disease Diagnosis}

Disease diagnosis is the process of determining which disease or condition explains a person's symptoms and signs~\cite{TP-Medical_diagnosis-web}. Early diagnosis wins time and money for patients~\cite{All-Ibd-web}. Nowadays, with the development of techniques, the diagnosis method has improved greatly.
First, the development of genomics techniques makes genetic data play an important role in diagnosis. For example, gene mutations are used to classify acute myelocytic leukemia (AML)~\cite{papaemmanuil2016genomic,bullinger2017genomics}, and two gene mutations have been included in the classification of myeloid neoplasms and AML by WHO~\cite{arber20162016}.
Second, electronic health records (EHRs) have increased dramatically recently. For example, 75.5\% of US hospitals had a basic EHR system~\cite{charles2013adoption} by 2014. Based on EHRs, Rajkomar et al.~\cite{rajkomar2018scalable} propose a fast healthcare interoperability resources (FHIR) format, making it possible to make the most of EHRs including free-text notes.
 In the following, three common diseases that Alzheimer's disease (AD)---age-related neurodegenerative disease~\cite{mckhann1984clinical}, Acute lymphoblastic leukemia (ALL)---the most common cancer in children~\cite{leukemia-web} and breast cancer---the most common diseases among women are discussed in detail.

\subsubsection{Representative disease diagnosis}

\textbf{Alzheimer's disease (AD) diagnosis}

AD is the most common age-related neurodegenerative disease resulting in an irreversible loss of memory and other cognitive functions in elderly people worldwide~\cite{mckhann1984clinical}. In 2006, the number of individuals with AD is 26.6 million~\cite{brookmeyer2007forecasting}. By 2050, this number will quadruple, by which time, 1 in 85 persons worldwide will be living with AD~\cite{brookmeyer2007forecasting}.  According to the order of severity, patients are classified as normal, mild cognitive impairment (MCI) or AD~\cite{adni-web}. Medical imaging of brains are usually used to diagnose AD, which is time consuming if the work is done manually.
 To tackle the problem, many automatic diagnostic systems have been developed. Among them, convolutional network (CNN) is the most prevalent  method and has good performance~\cite{litjens2017survey}.

\textbf{Acute lymphoblastic leukemia (ALL) diagnosis}

ALL is the most common cancer in children~\cite{leukemia-web} with a peak incidence at 2-5 years old~\cite{All-Ibd-web}. Without timely treatment, children with this serious blood pathology will die in a few weeks \cite{All-Ibd-web}. Early diagnosis helps provide timely and proper treatment for patients~\cite{All-Ibd-web}. Microscopic examination of blood or bone marrow smears is the only effective way to leukemia diagnosis~\cite{karthikeyan2017micros}. Generally, the method can be tackled by a classic sequence of steps: (1) enhancing image, (2) identification of white cells, (3) feature extraction, (4) classification~\cite{All-Ibd-web}.
 Besides image analysis, genomics studies have been introduced to inform disease classification in recent years ~\cite{bullinger2017genomics}.

\textbf{Breast cancer diagnosis}
Breast cancer has become one of the most common diseases among women that leads to death. Breast cancer can be diagnosed by classifying tumors. There are two different types of tumors, such as malignant and benign tumors. Doctors need a reliable diagnostic procedure to distinguish between these tumors~\cite{litjens2017survey}. Generally, it is very difficult to distinguish tumors even by the experts. Therefore, an automatic diagnostic system is needed to diagnose the tumors. The detection of breast cancer consists of three subtasks~\cite{gayathri2013breast}: (1) detection and classification of mass-like lesions, (2) detection and classification of micro-calcifications, (3) breast cancer risk scoring of images. By using and extending the results from the fields of machine learning, statistics, image processing and optimization, highly accurate diagnosis of breast is expected to be done even by untrained users.

\subsubsection{Publicly available data sets}

\textbf{Alzheimer's disease (AD) diagnosis}

Different materials such as clinical, cognitive, imaging, genetic, and biochemical biomarkers can all be used to define the progression of AD, and several researches try to determine the relationships between those data~\cite{adni-web}. ADNI and BioFINDER are two longitudinal studies for AD, which provide comprehensive data set of AD and have been used widely by researchers. However, several researches indicate that the data is still inadequate for solving the real problem. For example, the AD DREAM Challenge~\cite{allen2016crowdsourced} aims to benchmark  state-of-the-art algorithms in predicting AD based on publicly genetic and imaging data. However, the result is not so perfect,  and one possible reason is that the data used to train model is inadequate.

As a longitudinal multi-center study designed to develop clinical, imaging, genetic, and biochemical biomarkers for the early detection and tracking of AD, the Alzheimer's Disease Neuroimaging Initiative (ADNI)~\cite{adni-web} provides comprehensive data for AD. Since its foundation in 2003, it has made major contributions to AD research. Now it contains 483 subjects diagnosed with elderly control, 1001 subjects with MCI, and 437 subjects with AD. ADNI researchers collect several types of data: clinical data, genetic data, imaging data, and biospecimen data. Clinical dataset comprises recruitment, demographics, physical examinations, and cognitive assessment data saved as comma separated values (CSV) files. Genetic data contains genotyping and sequencing data. Images such as magnetic resonance imaging (MRI)  and positron emission tomography (PET) are available. Biospecimens includes blood, urine, and cerebrospinal fluid (CSF).

Swedish Biomarkers For Identifying Neurodegenerative Disorders Early and Reliably (BioFINDER)~\cite{biofinder-web} is another longitudinal study,  aiming to develop methods for early and accurate diagnosis of AD and Parkinson's disease (PD). It comprises more than 1600 subjects which undergo examinations of advanced MRI, CSF and plasma analysis, amyloid and tau PET,  clinical assessments and neuropsychological examinations~\cite{biofinder-web}.

\begin{table*}[h!]
  \begin{center}
    \caption{Data set for AD diagnosis}
    \label{tab:ad-data}
    \begin{tabular}{lp{3cm}p{3cm}p{2cm}p{1.5cm}p{1cm}p{1cm}}
     \toprule
      \textbf{Name} & \textbf{Task} & \textbf{Instances} & \textbf{Types}  & \textbf{Modality} & \textbf{Year} \\
      \midrule
      ADNI & Detect AD at early stage & 483 EC subjects, 1001 MCI subjects, 437 AD subjects & Clinical, genetic, imaging, biospecimen & MRI, PET, TXT & 2003-2018\\
      BioFINDER & Detect AD and PD at early stage & more than 1600 subjects & Clinical, genetic, imaging, biospecimen & MRI, PET, TXT & 2013-2018\\
      \bottomrule
    \end{tabular}
  \end{center}
  \begin{tablenotes}
      \small
      \item Abbreviation: AD = Alzheimer disease; EC = elderly controls; MCI = mild cognitive impairment; PD =  Parkinson's disease.
    \end{tablenotes}
\end{table*}

\textbf{Acute lymphoblastic leukemia (ALL) diagnosis}

There are many comprehensive dataset for caners (e.g. TCGA). But as a branch of cancer, the data for ALL is relatively decentralized. Blood or bone marrow smears are key materials to diagnose ALL~\cite{karthikeyan2017micros}.  ALL-IDB is a public image database for ALL that has been studied widely. However, several researchers~\cite{mahapatra2016retinal,orlando2017convolutional} claim that hundreds of images are not enough to build a robust CNN, so more public data is still needed for ALL diagnosis. Besides, genomic information is also provided for ALL such as TARGET~\cite{leukemia-web} and BioGPS~\cite{goto-welcome-web}.

Acute lymphoblastic leukemia image database (ALL-IDB)~\cite{All-Ibd-web} is a public dataset of microscopic images of blood samples, based on which researchers can evaluate their segmentation and classification algorithms. The format of the images is JPG with 24 bit color depth and resolution 2592 x 1944. ALL-IDB1 contains 108 images of 39000 blood elements, where lymphocytes are manually labeled by experts. Cropped areas of interest of cells belonging to ALL-IDB1 are collected as ALL-IDB2 dataset.

Therapeutically applicable research to generate effective treatments (TARGET)~\cite{leukemia-web}, a project of national institutes of health (NIH), determines molecular changes that drive childhood cancers by genomic approaches. The ALL pilot phase (Phase I) has produced genomic profiles of nearly 200 B-cell ALL patient cases for molecular alterations. Nucleic acid samples data, extracted from peripheral blood and bone marrow tissues,
is included in each fully-characterized case.
 The dataset consists of clinical information, tissue pathology data, chromosome-specific copy number alterations,  sequence data of single amplicons, and mutations. BioGPS~\cite{goto-welcome-web} is a gene annotation portal, which supplies serval ALL genetic dataset, the biggest of which contains 207 samples provided by children's oncology group (COG) study P9906 for high-risk pediatric ALL~\cite{willman2013identification}. For each subject the dataset provides genome structure information such as BCR-ABL, E2A-PBX1, TEL-AML and clinical information such as central nervous system (CNS) status, white blood cell (WBC), age, gender and etc.

\begin{table*}[h!]
  \begin{center}
    \caption{Data set for ALL diagnosis}
    \label{tab:all-data}
    \begin{tabular}{lp{4cm}p{2cm}p{1cm}p{2cm}p{1cm}p{1cm}}
     \toprule
      \textbf{Name} & \textbf{Task} & \textbf{Instances} & \textbf{Types}  & \textbf{Modality} & \textbf{Year} \\
      \midrule
      TARGET Phase I & Determine molecular changes that drive childhood cancers & 200 subjects & Clinical, genetic & TXT & 2009-2018\\
      BioGPS COG P9906  &  Evaluate a regimen in patients with high risk B-precursor ALL & 267 subjects & Clinical, genetic & TXT & 2011\\
      ALL-IDB & Evaluate the algorithms for image segmentation and classification & 108 images & Imaging & JPG & 2010\\
      \bottomrule
    \end{tabular}
  \end{center}
\end{table*}

\textbf{Breast Cancer Diagnosis}

Digital imaging databases are needed for mammographic image analysis research. For the sake of accurate labeling of images, free-text report databases are necessary as well, which can be leveraged to turn the reports into accurate annotations automatically for network training~\cite{litjens2017survey}. Besides, it is recommendable to use gene expression databases  to be able to connecting cancer phenotypes to genotypes.

So far, all three types of breast databases mentioned above have been developed. Unfortunately, most of the large databases are not publicly available and many stale databases are still in use.

INbreast, mini-MIAS, DDSM, BCDR-FMR, Breast Cancer Wisconsin Dataset and TCGA-BRCA are the most frequently-used mammographic mass classification datasets. These databases do represent a constructive and practical contribution to computer vision research in mammography (in short, MG) and it is expected that they will encourage the production of more extensive collections of data. Table \ref{tab2333} shows the most commonly-used datasets for diagnosing breast cancer.

\begin{table}[htbp]
\caption{Overview of datasets for breast cancer diagnosis. MG stands for  mammography.}
\begin{center}
\scalebox{0.78}[0.78]{
\renewcommand{\arraystretch}{1.5}
\begin{tabular}{|p{0.32in}|p{1.1in}|p{0.8in}|p{0.6in}|p{0.35in}|}
\hline
\textbf{Name} & \textbf{Task} & \textbf{Instance} & \textbf{Types} & \textbf{Modality}  \\ \hline
INbreast & develop breast cancer CAD systems & 115 cases , 410 images & imaging & MG \\ \hline
MIAS & mammographic image analysis research & 322 images & imaging & MG \\ \hline
DDSM & mammographic image analysis research & 2620 cases, 43 volumes & imaging, expert ground-truth, metadata & MG \\ \hline
BCDR-FMR & lesion classification & 1010 cases, 3703 images & imaging, metadata & MG \\ \hline
BCW & lesion classification & 569 instances, 32 attributes & imaging, metadata & MG \\ \hline
TCGA-BRCA & connecting cancer phenotypes to genotypes & 139 cases, 230167 images & imaging, clinical and genomic data & MR, MG \\ \hline
\end{tabular}
}
\label{tab2333}
\end{center}
\end{table}

The INbreast database contains 115 examples, which includes 90 examples from breast-affected women and 25 examples from mammectomy women~\cite{moreira2012inbreast}. The INbreast database includes many types of lesions~\cite{moreira2012inbreast}. The comparative advantages of the INbreast database are its huge amount of examples together with accurate labels~\cite{moreira2012inbreast}. It is glad to see that this database can strongly support the future work on breast cancer diagnosis~\cite{moreira2012inbreast}.

The Mammographic Image Analysis Society (MIAS) is a digital mammography (in short, MG) dataset. It includes 322 digital images,  and contains both abnormal examples and normal examples\cite{suckling1994mammographic} . The entire database, when compressed, occupies less than 2 GBytes fitting onto a single 8 mm magnetic tape. Copies are available for research purposes. The mini-MIAS database is available for scientific research at no cost, provided that they must abide by the licence agreement when using the imagery.

The Digital Database for Screening Mammography (DDSM) contains digitized mammograms together with related labels and other detailed information~\cite{rose2006web}. The DDSM database is freely available through the website~\cite{rose2006web}.

The Breast Cancer Digital Repository (BCDR-FMR) is a comprehensive labeled dataset, which provides digital content (digitized film mammography images) and associated metadata (clinical history, segmented lesions BI-RADS classified, image-based descriptors, biopsy proven, etc.). The BCDR-FMR establishes a novel reference to develop breast cancer diagnosis methods~\cite{lopez2012bcdr}.

The Breast Cancer Wisconsin (Diagnostic) Dataset is comparatively abundant in examples, including 569 patients, and for each instances, there are 32 attributes to describe it and ten features to measure it. There are both qualitative and quantitative features in the dataset. All feature values are recorded with four significant digits. This dataset consists of 212 malignant cases and 357 benign cases~\cite{maddix2014diagnosing}.

Nowadays, tumor gene expression analytical techniques based on DNA microarray have been applied to diagnose breast cancer~\cite{ramaswamy2001multiclass}. TCGA-BRCA project has explored the most comprehensive gene expression database. However, analytical algorithms, which can solve gene expression-based diagnosis problems, have yet to be established~\cite{ramaswamy2001multiclass}.

Traditional PACS (Picture archiving and communication)  systems preserve structured reports described by radiologists~\cite{moreira2012inbreast}. To optimally leverage free-text reports to train network, we can automatically turn these reports into precise labels or structured annotations~\cite{moreira2012inbreast}. However, most PACS databases are unavailable.





\subsection{Drug discovery}

Drug discovery is the process of finding drug candidates that can be used as new drugs~\cite{drews2000drug}.
The motivation for drug discovery is because there are no suitable medical products for certain diseases~\cite{hughes2011principles}.
Drug discovery is generally divided into the following steps: target identification and validation, screening and lead discovery, lead optimization and retrosynthetic analysis~\cite{sliwoski2014computational, drews2000drug}.
Despite advances in biotechnology, drug discovery remains an expensive, difficult and inefficient process~\cite{kolb2003growing}.

In target identification and validation phase, we need to determine the pathogenic factors of the disease, from DNA to RNA to protein characterization~\cite{chen2016leveraging}.

Lead discovery and drug screening refer to the process of assessing the biological activity, pharmacological effects, and medicinal value of a substance that may be used as a drug using an appropriate method~\cite{ramsundar2015massively}.
Drug screening mainly includes high-throughput screening (HTS) and virtual drug screening~\cite{hughes2011principles}.
HTS methods can be performed by robots at the same time for millions of tests, so the cost is very high.
Because real drug screening requires the construction of large-scale compound libraries, extraction or cultivation of a large number of target enzymes or target cells necessary for experiments, and the need for complex device support, Drug screening requires a huge investment.
The virtual drug screening method simulates the process of drug screening on a computer, predicts the possible activity of the compound, and then performs targeted physical screening of compounds that are likely to become drugs.

The final stage of drug discovery is lead optimization and retrosynthetic analysis.
The purpose is to maintain the advantageous properties of the lead while improving the defects of the lead structure.
Retrosynthetic analysis is to generate a synthetic route for a given target molecule~\cite{nicolaou1999total}.
Retrosynthetic analysis is to give a desired target molecule, using molecular compounds that can be directly synthesized, to give several possible synthetic routes of the target molecule~\cite{segler2018planning}.
The difficulty is that if the molecular compound to be synthesized is very complicated, the chemist may have to consult a lot of relevant literature, and also carry out repeated practice analysis in order to finally obtain several possible synthetic routes~\cite{nicolaou2005total}.
However, chemists may not be able to find a reasonable synthetic route because of knowledge or time constraints, or only find a few routes.

Many types of data are needed for drug discovery.
At the stages of  target identification and validation as well as lead optimization and retrosynthetic analysis,
gene expression data and molecular-level data are needed, including compound structure, properties, and related chemical reactions.
In drug screening, we need a drug sensitivity database and a toxicity database~\cite{chen2016leveraging}.
Several databases related to drug discovery are listed in Table~\ref{tabdrugdisc}.

\begin{table}[htbp]
    \caption{Overview of datasets for drug discovery}
    \begin{center}
    \scalebox{0.78}[0.78]{
    \begin{tabular}{|p{0.5in}|p{1in}|p{1in}|p{1in}|}
    \hline
    \textbf{Name} & \textbf{Task} & \textbf{Instance} & \textbf{Types}  \\ \hline
    Reaxys & discover chemical structures, properties and reactions & more than 28 million responses, more than 18 million substances, and more than 4 million documents & relevant literature, precise compound properties and reaction data \\ \hline
    SIOC & for chemical research & - & compound structure and identification, natural products and pharmaceutical chemistry, chemical literature, chemical reactions and comprehensive information \\ \hline
    PubChem  BioAssay & deliver free and easy access to all deposited data, and to provide intuitive data analysis tools~\cite{wang2011pubchem} & 500,000 descriptions of assay protocols, covering 5000 protein targets, 30,000 gene targets and providing over 130 million bioactivity outcomes~\cite{wang2011pubchem} & chemical structure and biological properties of small molecules and RNAi reagents~\cite{wang2011pubchem} \\ \hline
    TCM & construct the first traditional Chinese medicine database for molecular docking simulation~\cite{chen2011tcm} & 20,000 pure compounds isolated from 453 TCM components~\cite{chen2011tcm} & molecular attributes, substructures, TCM components, and TCM classifications \\ \hline
    ChEMBL & address a wide range of drug discovery problems & 2,275,906 compound records, 12,091 targets & compound structure, biological or physicochemical measurements of these compounds and information on the goals of these assays arerecorded in a structured form~\cite{bento2014chembl} \\ \hline
    \end{tabular}}
    \label{tabdrugdisc}
    \end{center}
\end{table}

\subsubsection{Gene expression database and molecular-level database}
The Reaxys database is produced by Elsevier, and is a rich database of chemical values and facts.
Reaxys integrates the contents of Beilstein, Patent and Gmelin into an unified resource that includes more than 28 million responses, more than 18 million substances, and more than 4 million documents.
It helps users identify promising new projects, terminate ineffective lead compounds, and design economical and high-yield synthetic routes that maximize time and cost savings.

The Shanghai Institute of Organic Chemistry's~(SIOC) database~(\url{http://www.organchem.csdb.cn}) group is a comprehensive information system for chemical research and development.
It provides compound structure and identification, natural products and pharmaceutical chemistry, chemical literature, chemical reactions and comprehensive information.
Chemical reaction condition retrieval is to search for matching chemical reactions in the database by chemical reaction conditions such as reactants, products, catalysts, solvents, reagents, and the like. The user can search for the relevant reaction by the reactant, the English name of the product, the reaction conditions, the catalyst, etc.

The Taiwan traditional Chinese medicine (TCM) database~\cite{chen2011tcm} is currently the largest non-commercial TCM database in the world.
This web-based database contains more than 20,000 pure compounds isolated from 453 TCM components~\cite{chen2011tcm}.
All data are easily accessible to all researchers. In the past eight years, many volunteers have spent time in analyzing Chinese medicine ingredients in the Chinese medical literature and building structural files for each of the isolated compounds.

\subsubsection{Lead discovery and drug screening database}
PubChem BioAssay database~\cite{wang2011pubchem} is a public resource for archiving the chemical structure and biological properties of small molecules and RNAi reagents.
The PubChem BioAssay database currently includes bioactivity from high-throughput screening and medicinal chemistry studies~\cite{wang2011pubchem}.
In addition, the PubChem BioAssay database contains dozens of high-throughput RNAi screens for complete genomes.
These data, combined with other NCBI resources, make PubChem a public information system widely used in chemical biology and drug discovery research~\cite{wang2011pubchem}.


ChEMBL~\cite{bento2014chembl} is an open, large-scale bioactivity database containing information manually extracted from the medicinal chemistry literature~(\url{https://www.ebi.ac.uk/chembl}).
The ChEMBL database currently contains information extracted from more than 51,000 publications, as well as bioactive data sets from 18 other databases~\cite{bento2014chembl}. The data mainly includes screening results and bioactivity data.

 \section{Do state-of-the-art and state-of-the-practise algorithms perform a good job?}
 ~\label{sec:algorithm}

\subsection{Quantified Self}
With wearable devices deployed in recent years, more and more physiological and functional data is captured continuously for healthcare applications~\cite{swan2013quantified}. In the previous work, traditional statistical methods are widely used. There are two challenges and  opportunities for handling QS data. On one hand, although deep learning has been applied in high performance platforms successfully, it does not perform well on low-power wearable devices due to resource limits~\cite{chen2015deep}. On the other hand,
 sensor data of QS is mostly time-series~\cite{swan2013quantified}.

In the previous work, the data are usually analysed by traditional technologies like linear regression or other statistical methods. For example, Angeles et al.~\cite{angeles2016wearable} use statistical algorithms to distinguish between non-mimicked and mimicked tests for all the Parkinson's primary symptoms, with very convincing differences. To evaluate the patients in rehabilitation recovery progress~\cite{ravi2016deep}, Chen et al.~\cite{chen2015wearable} use some validation techniques, such as 10-fold cross-validation. This technique can classify the types of exercise and determine if their postures are appropriate. The overall accuracy of posture recognition is 88.26\% and that of type classification is 97.29\%~\cite{chen2015wearable}. It is believed to be beneficial for patients to effectively rehabilitate.

We summarize state-of-the-practise and state-of-the-art work of applying machine learning for QS in Table~\ref{tab511}.
To identify common daily living activities for chronic disease management, Atallah et al.~\cite{atallah2009real} use a two-stage Bayesian classifier to evaluate the condition of patients with chronic obstructive pulmonary disease (COPD). The development of Bayesian classification framework can explain the errors in sensor data, and classification accuracy of different activities~\cite{atallah2009real}.
Methods like SVM (support vector machines) and decision trees are trained to classify the data~\cite{lu2017towards, ignatov2016human, walse2016study}.
For human physical activity recognition, Ignatov et al.~\cite{ignatov2016human} propose a method using k-nearest neighbor and DNN as an alternative to process time-series data. Their method has high accuracy. When using a set of segmentation and KNN, it achieves nearly 96\% recognition accuracy~\cite{ignatov2016human}. 
To recognize activity, Catal et al.~\cite{catal2015use} propose a method that combines multiple classification methods such as J48 decision tree, Multi-Layer Perceptrons (MLP) and Logistic Regression techniques.
Using deep learning methods, including CNN (convolutional neural networks), RBM (Restricted Boltzmann machines), and DBN (deep belief networks), from the input data, the machine can directly learn a set of features which are discriminative~\cite{zeng2014convolutional, yang2015deep, chen2015deep}. Alsheikh et al.~\cite{alsheikh2016deep} use a method based on DBNs and RBMs. The method uses multiple hidden layers to recognize activity. This work proves that these models have better recognition accuracy of human activities, using a large number of unlabeled acceleration samples to extract unsupervised features and avoid expensive manual features design in existing systems~\cite{alsheikh2016deep}.

Nowadays, the information of the data requires efficient ways of classification and analysis if deep learning is a good choice in some cases~\cite{ravi2016deep}. Deep learning is a promising technique that could extract information and infer from big data by using multiple processing layers~\cite{lecun2015deep}. 


\begin{table}[htbp]
\caption{The summary of algorithms used in QS.}
\begin{center}
\scalebox{0.78}[0.78]{
\renewcommand{\arraystretch}{1.5}
\begin{tabular}{|p{0.4in}|p{0.7in}|p{0.7in}|p{0.7in}|p{0.5in}|p{0.2in}|}
\hline
\textbf{Reference} & \textbf{Method} & \textbf{Modality} & \textbf{Application} & \textbf{Database} & \textbf{Year} \\ \hline
~\cite{lu2017towards} & GMM, HC, K-means, K-medoids, SC & time series data & HAR & none & 2017 \\ \hline
~\cite{angeles2016wearable} & statistical & time series data & Quantify Parkinsonian symptoms & none & 2016 \\ \hline
~\cite{ignatov2016human} & LR, NN, SVM, J48, KNN & time series data & HAR & none & 2016 \\ \hline
~\cite{walse2016study} & adaboost & time series data & HAR & WISDM & 2016 \\ \hline
~\cite{alsheikh2016deep} & DBN and RBM & time series data & HAR & WISDM, Daphnet, Skoda & 2016 \\ \hline
~\cite{chen2015wearable} & cross-validation & time series data & Rehabilitation Exercise Assessment for Knee Osteoarthritis & none & 2016 \\ \hline
~\cite{catal2015use} & ensemble of classifiers & time series data & HAR & WISDM & 2015 \\ \hline
~\cite{yang2015deep} & DCNN & time series data & HAR & Opp, Hand Gesture & 2015 \\ \hline
~\cite{chen2015deep} & CNN & time series data & HAR & none & 2015 \\ \hline
~\cite{zeng2014convolutional} & CNN & time series data & HAR & Opp, Skoda, Actitracker & 2014 \\ \hline
~\cite{atallah2009real} & Bayesian classifier & time series data & chronic obstructive pulmonary disease & none & 2009 \\ \hline
\end{tabular}
}
\label{tab511}
\end{center}
\begin{tablenotes}
      \small
      \item \textit{\textbf{Abbreviation:}} GMM = Gaussian Mixture Model with Expectation-Maximization; HC = a hierarchical method; SC = spectral clustering.
\end{tablenotes}
\end{table}

\subsection{ Disease classification}
Disease classification,  which groups diseases together based on their similarities,  is expected to promote understanding and curing  human diseases~\cite{national2011toward}. The conventional methods  classify human diseases through 4 dimensions containing pathogen, the original component causing diseases, pathology and clinical patient behavior. However, with the improvement of human gene analysis and molecular biology, researchers obtain a large amount of relative information. And with the molecular level information, our knowledge network of diseases has been refreshed and our cognition of diseases has also been improved. Consequently, these two significant changes provide an appropriate opportunity for researchers to develop more precise methods to classify human diseases from the molecular level.

Table~\ref{tab:table2} shows a variety of mature algorithms relating to classification and clustering.
Though, at present many algorithms of disease taxonomy has been developed, researchers need to develop more effective and robust algorithms to meet the needs of disease classification with the increase of different type data of molecular biology.

\begin{table*}[h!]
  \begin{center}
    \caption{Summary of algorithms for disease classification}
    \label{tab:table2}
    \begin{tabular}{|c|p{1.7in}|c|p{2in}|c|c|}
     \toprule
      \textbf{Reference} & \textbf{Method} & \textbf{Modality} & \textbf{Application}  & \textbf{Data} & \textbf{Year} \\
     \midrule
      \cite{PMID:29625048}&\tabincell{c}{iCluster~\cite{Shen2009Integrative}}&genomic&\tabincell{c}{Clustering cancer from chromosome, \\DNA, mRNA and protein level.}&TCGA&2018\\
     \midrule
      \cite{Dunne2017Cancer}&\tabincell{c}{random forest, DIANA clustering,\\two-dimensional hierarchical \\clustering}&genomic&\tabincell{c}{Classifying colorectal cancer with gene \\information of cancer cells.}&GEO&2017\\
     \midrule
      \cite{bailey2016genomic}&\tabincell{c}{non-negative matrix factorization}&genomic&\tabincell{c}{Clustering  pancreatic cancer information of \\recurring altering genes.}&APGI&2016\\
            \midrule
      \cite{Guinney2015The}&\tabincell{c}{Markov cluster (MCL) algorithm}&genomic&\tabincell{c}{Reclassifying 6 classification system of \\Colorectal cancer into 4 consensus \\molecular subtypes.}
&\tabincell{c}{CCLE, \\GSK, \\ Sanger}&2015\\
      \midrule
     \cite{hoadley2014multiplatform}&\tabincell{c}{cluster-of -cluster \\assignments(COCA) algorithm}&genomic&\tabincell{c}{Clustering 12 cancer types from gene and \\protein level to find new subtypes of \\the cancer.}&TCGA&2014\\
      \midrule
     \cite{Shen2012Integrative}&iCluster~\cite{Shen2009Integrative}&genomic&\tabincell{c}{Cancer subtype classification and discovery.}&TCGA&2012\\

      \bottomrule
    \end{tabular}
  \end{center}
\end{table*}

Hoadley et al.~\cite{hoadley2014multiplatform} reclassified 12 kinds of cancer from molecular level, which originate from different organs. In the research, the cluster-of-cluster assignments (COCA) algorithm, which is a kind of agglomerative hierarchical clustering method using Pearson correlation as the distance measurement, is adopted to cluster the genomic and proteomic data from different platforms. According to the research result, the 12 cancers conventionally classified according to their original organs are uniformly reclassified into 11 major subtypes with their causes of gene and protein, which provides a new idea to the cancer clinical treatment strategy.

Hoadley at al.~\cite{hoadley2018cell} conduct a molecular reclassification of 33 cancer types from the TCGA platform. During the process of the reclassification, researchers utilize iCluster algorithm, a joint latent variable model-based clustering algorithm to deal with different type data, to cluster molecular data involving chromosome, DNA, mRNA, miRNA and protein from TCGA. Based on the consequence of the study, researchers re-cluster 33 cancer types into 28 cancer subtypes. Besides, they discover and verify the dominated position of cell-of-origin pattern in cancer molecular classification. The molecular similarities of cancer subtypes contribute to the improvement of future therapy of cancer~\cite{hoadley2018cell}.

Bailey at al.~\cite{bailey2016genomic} identify various subtypes of pancreatic ductal adenocarcinomas through genetic analysis. The non-negative matrix factorization clustering, which uses the matrix factorization technology to cluster samples, is exploited to cluster recurring altering genes of 456 pancreatic cancer patients. This study discovers 4 subtypes of pancreatic cancer with respectively specific pathological characteristics, which provides beneficial information for the inference of the development of pancreatic cancer and a new idea of clinically therapeutic strategy of pancreatic cancer~\cite{bailey2016genomic}.

Guinney et al.~\cite{Guinney2015The} construct a consensus classification of colorectal cancer (CRC), since the translational and clinical utility of gene expression-based subtyping is hampered by discrepant results from different researches. Firstly, the Markov cluster (MCL) algorithm is applied to obtain the consensus molecular subtypes (CMS) from 6 CRC classification systems. And then a random forest model is utilized to classify new samples into CMS. The research classifies CRC into 4 consensus molecular subtypes microsatellite instability immune, canonical, metabolic and mesenchymal~\cite{Guinney2015The}. Each subtype has the clear biological interpretability, which is very important in the treatment of patients.

\subsection{Disease diagnosis}
\subsubsection{Alzheimer's disease (AD) diagnosis}
Depending on severity, patients are usually diagnosed as normal controls (NC), MCI or AD. Different materials such as neuroimaging, biospecimen and genetic data are used to diagnose AD. In the previous work, deep learning techniques are extensively used to analyze medical imaging by many researchers and have achieved good performance. Although some studies ~\cite{hansson2018csf, lorenzi2018susceptibility} show that biospecimens and genetic data provide alternatives to neuroimaging in AD diagnosis, those data is scarcely used to diagnose AD in practice. So researchers should pay more attention to develop comprehensive  and integrative diagnosis method based on different materials, which is likely to achieve much higher precision.

Analyzing neuroimagings such as MRI and PET images is a prevalent method for  AD diagnosis. Jie et al.~\cite{jie2015manifold} propose a manifold regularized multitask feature learning method and then classify patients to three categories: AD, MCI and NC. The algorithm reaches 95.03\% accuracy tested on MRI and PET images from ADNI, which contains 202 subjects: 51 AD patients, 99 MCI patients, and 52 NC. Suk et al.~\cite{suk2016deep} utilize sparse regression models as target-level representation learner and build a deep convolutional neural network for AD identification. The algorithm reaches 90.28\% accuracy on a baseline MRI dataset of 805 subjects, including 186 AD, 393 MCI, and 226 NC, from the ADNI database. Shi et al. ~\cite{shi2017nonlinear} use thin-plate spline (TPS) based nonlinear feature transformation and stack denoising sparse auto-encoder (DSAE) deep fusion for AD staging analysis, and the approach reaches 91.95\% accuracy. The experiment is performed on a sub dataset of MRIs together with their whole brain masks selected from ADNI, which contains 338 subjects: 94 patients with AD, 121 with MCI and 123 NCs. Shi et al.~\cite{shi2018multimodal} develop a multi-modal stacked deep polynomial networks  (MM-SDPN) algorithm  to fuse and learn feature representation from multi-modal neuroimaging data for AD diagnosis,  and the approaches reach 97.13\% accuracy. Data from ANDI are used here, consisting of MRI and PET images from 202 subjects: 51 AD patients, 99 MCI patients, and 52 NC. The performances of different classifiers are summarized in Table \ref{tab:ad-acc}.

Several studies indicate that CSF biomarkers provide alternatives to MRI and PET images in AD diagnosis. Hansson et al.~\cite{hansson2018csf} indicate that tau/$A\beta$ ratios are as accurate as semiquantitative PET image assessment. Mattsson et al.~\cite{mattsson2018comparing} provide  evidence that when identifying early AD, CSF tau and  ${}^{18}$F-AV-1451 PET have similar performance but MRI measures have lower area under the receiver operating characteristic curve (AUROC). Besides, they find when identifying  mild to moderate AD, ${}^{18}$F-AV-1451 PET is superior to CSF tau.

Although genetic information is provided in ADNI, the studies are limited mainly because of the nontrivial validation of correlations between genetic variants and phenotype. However several studies show that certain genes are related to AD. For instance, Lorenzi et al.~\cite{lorenzi2018susceptibility} identify a link between tribbles pseudokinase 3 (TRIB3) and the stereotypical pattern of gray matter loss in AD.

\begin{table}[h]
  \begin{center}
    \caption{Classification result of different algorithms for AD diagnosis}
    \label{tab:ad-acc}
    \begin{tabular}{|c|c|c|c|c|c|c|}
     \toprule
     \textbf{Reference} & \multicolumn{3}{c|}{AD vs. NC} & \multicolumn{3}{c|}{MCI vs. NC} \\
      \textbf{} & \textbf{ACC} & \textbf{SEN} & \textbf{SPE}  & \textbf{ACC} & \textbf{SEN} & \textbf{SPE} \\
      \midrule
      \cite{shi2018multimodal} & 97.13 &95.93 & 98.53 &	87.24 &	97.91 &	67.04\\
      \cite{shi2017nonlinear} & 91.95 &	89.49 &	93.82 &	83.72 &	84.74 &	82.72 \\
     \cite{suk2016deep}  & 90.28 &	92.65 &	89.05 &	74.20 &	78.74 &	66.30 \\
     \cite{jie2015manifold} & 95.03 &	94.90 &	95.00 &	79.27 &	85.86 &	66.54 \\

      \bottomrule
    \end{tabular}
  \end{center}
  \begin{tablenotes}
      \small
      \item Abbreviation: AD = Alzheimer disease; MCI = mild cognitive impairment; NC = normal controls; ACC = Accuracy; SEN = Sensitivity; SPE = specificity.
    \end{tablenotes}
\end{table}

\begin{table*}[h!]
  \begin{center}
    \caption{Overview of papers for AD diagnosis}
    \label{tab:ad-algorithm}
    \begin{tabular}{lp{2cm}p{2cm}p{3cm}p{2cm}p{1cm}p{1cm}}
     \toprule
      \textbf{Reference} & \textbf{Method} & \textbf{Modality} & \textbf{Application}  & \textbf{Data} & \textbf{Year} \\
      \midrule

     \cite{shi2018multimodal} & MM-SDPN & MRI, PET & AD diagnosis &	ADNI &	2018\\
     \cite{hansson2018csf}  & Comparation&	CSF, PET &  Clinical progression prediction &	ADNI, BioFINDER & 2018 \\
     \cite{mattsson2018comparing} & Comparation & CSF, PET & AD diagnosis  &	BioFINDER &	2018 \\
     \cite{lorenzi2018susceptibility} & Statistic & Genetic, PET & Genetic underpinnings of  AD & ADNI &	2018\\
     \cite{shi2017nonlinear} & DSAE & MRI & AD diagnosis &	ADNI &	2017 \\
     \cite{suk2016deep}  & CNN &	MRI & AD diagnosis &	ADNI &	2016 \\
     \cite{jie2015manifold} & Laplacian regularizer & MRI, PET, CSF & AD diagnosis & ADNI & 2015\\

      \bottomrule
    \end{tabular}
  \end{center}
  \begin{tablenotes}
      \small
      \item Abbreviation: CNN = convolutional neural network; DSAE = stacked denoising sparse auto-encoder; MM-SDPN = multimodal stacked deep polynomial networks.
    \end{tablenotes}
\end{table*}

\subsubsection{Acute lymphoblastic leukemia (ALL) diagnosis}
Microscopic examination of blood or bone marrow smears is the only effective way to leukemia diagnosis ~\cite{karthikeyan2017micros}.  Several methods ~\cite{mohapatra2014ensemble,amin2015recognition,karthikeyan2017micros,abdeldaim2018computer} have been provided to classify cells to be cancerous or noncancerous, and the  most achieve the accuracy more than 95\%. For instance, Abdeldaim et al.~\cite{abdeldaim2018computer} present a computer-aided ALL diagnosis system, which first segments each cell in the microscopic images, and then classifies each segmented cell to be normal or affected. The experiment based on ALL-IDB2 achieves the accuracy of 96.42\% with a KNN classifier. Gene data may also be used to diagnose ALL. For instance, by examining gene expression profiles, Willman et al.~\cite{willman2013identification} find that the clusters are associated with either specific clinical features or treatment response characteristics in the children with high risk B-precursor ALL.

\begin{table*}[h!]
  \begin{center}
    \caption{Overview of papers for ALL diagnosis}
    \label{tab:all-algorithm}
    \begin{tabular}{lp{2cm}p{2cm}p{3cm}p{2cm}p{1cm}p{1cm}}
     \toprule
      \textbf{Reference} & \textbf{Method} & \textbf{Modality} & \textbf{Application}  & \textbf{Data} & \textbf{Year} \\
      \midrule
      \cite{abdeldaim2018computer}  & K-NN &	JPG & Image segmentation &	ALL-IDB &	2018 \\
      \cite{karthikeyan2017micros} & Fuzzy C Means & JPG  & Image segmentation & Taken form Google & 2017\\
      \cite{amin2015recognition}  & K-Means &	JPG & ALL classification &	Isfahan Al-Zahra and Omid hospital &	2015 \\
     \cite{mohapatra2014ensemble}  & Ensemble classifier &	JPG & ALL classification & Ispat General Hospital & 2014 \\
    \cite{willman2013identification}  & Cluster & TXT & ALL feature selection &	TARGET &	2013 \\

      \bottomrule
    \end{tabular}
  \end{center}
\end{table*}

\subsubsection{Breast Cancer Diagnosis}
Machine learning approaches have been extensively used in the diagnosis of breast cancer. Researchers have focused on devising better algorithms to automate the detection of cancerous cells. Table~\ref{tab233} shows the state-of-the-art algorithms  diagnosing breast cancer.

\begin{table}[htbp]
\caption{Summary of breast cancer diagnosis.}
\begin{center}
\scalebox{0.78}[0.78]{
\renewcommand{\arraystretch}{1.5}
\begin{tabular}{|p{0.4in}|p{0.45in}|p{0.35in}|p{1.2in}|p{0.7in}|p{0.2in}|}
\hline
\textbf{Reference} & \textbf{Techinique} & \textbf{Modality} & \textbf{Application} & \textbf{DB} & \textbf{Year} \\ \hline
\cite{zhu2018adversarial} & GAN & MG & Mabss segmentation & INbreast, DDSM-BCRP & 2018 \\ \hline
\cite{zhu2017deep} & MIL-CNN & MG & Lesion classification & INbreast & 2017 \\ \hline
\cite{kooi2017discriminating} & CNN & MG & Lesion classification & INbreast, DDSM & 2017 \\ \hline
\cite{sun2017enhancing} & CNN & MG & Semi-supervised CNN for classification of masses & FFDM & 2017 \\ \hline
\cite{fotin2016detection} & CNN & MRI & Breast and fibro glandular tissue segmentation & Self-produced data & 2017 \\ \hline
\cite{wang2017detecting} & CNN & MG & Detection of cardiovascular disease based on vessel calcification & Self-produced data & 2017 \\ \hline
\cite{albarqouni2016aggnet} & M-CNN & H\&E & Mitosis detection & AMIDA & 2016 \\ \hline
\cite{cheng2016computer} & SAE & US, CT & Lesion classification & Self-produced data & 2016 \\ \hline
\cite{kallenberg2016unsupervised} & CAE & MG & Breast density segmentation, breast cancer risk scoring & Self-produced data & 2016 \\ \hline
\cite{zhang2016deep} & RBM & US & Lesion classification & Self-produced data & 2016 \\ \hline
\cite{kooi2017large} & CNN & TS & Mass detection & DDSM & 2016 \\ \hline
\cite{arevalo2016representation} & CNN & MG & Lesion classification & BCDR & 2016 \\ \hline
\cite{dubrovina2018computational} & CNN & MG & Tissue classification using regular CNNs & Self-produced data & 2016 \\ \hline
\cite{dhungel2016automated} & CNN & MG & Lesion classification & Inbreast & 2016 \\ \hline
\cite{paeng2017unified} & CNN & MG & Mass localization & TCGA & 2016 \\ \hline
\cite{huynh2016digital} & CNN & MG & Mass classification & Collected from University of Chicago Medical Center & 2016 \\ \hline
\cite{kisilev2016medical} & CNN & MG & Lesion classification & Self-produced data & 2016 \\ \hline
\cite{qiu2016initial} & CNN & MG & Cancer risk score & Self-produced data & 2016 \\ \hline
\cite{samala2016deep} & CNN & TS & Micro calcification detection & Collected from the University of Michigan & 2016 \\ \hline
\cite{samala2016mass} & CNN & TS & Transfer mammographic masses to tomosynthesis & Self-produced data & 2016 \\ \hline
\cite{akselrod2016region} & CNN & MG & Mass classification& Inbreast & 2016 \\ \hline
\cite{fonseca2015automatic} & CNN & MG & Lesion classification& MIAS & 2015 \\ \hline
\cite{jamieson2012breast} & ADN & MG,US & Mass classification & Self-produced data & 2012 \\ \hline

\end{tabular}
}
\label{tab233}
\end{center}
\begin{tablenotes}
      \small
      \item \textit{\textbf{Abbreviation:}} M-CNN= Multi-Stream CNN; MIL-CNN= Multi-instance Learning CNN; CAE= Convolutional Auto-Encoders; SAE= Stacked Auto-Encoders; H\&E= Hematoxylin \& Eosin Histology Images; MG= Mammography; US= Ultrasound; CT= Computed Tomography.
\end{tablenotes}
\end{table}

Huynh et al.~\cite{huynh2016digital} learn the features on mammography pictures by the use of CNN algorithm, and an SVM model is used to classify the derivative features into three categories, which include benign, cystic and malignant. They apply their algorithm on a dataset which include 607 breast images and procure an AUC (Area Under the Curve) of 86\%~\cite{huynh2016digital}.

Wang et al.~\cite{wang2017detecting} develop a 12-layer CNN for breast arterial calcification (BAC) detection. The micro-calcium detection and diagnosis algorithm yields 96.24\% accuracy, and the inferred micro-calcium lesion is close to the ground truth~\cite{wang2017detecting}.

Sun et al.~\cite{sun2017enhancing} present a semi-supervised learning method based on graph by the use of CNN to diagnose breast cancer. They obtain an AUC of 88.18\% on a dataset which contains both labeled and unlabeled data~\cite{sun2017enhancing}.

Kooi et al.~\cite{kooi2017discriminating} develop a computer-aided diagnosis technique to diagnose benign solitary lesions from malignant masses on digital mammogram. The algorithm obtains an AUC of 87\%, which is better than the other algorithms~\cite{kooi2017discriminating}.

Zhu et al.~\cite{zhu2017deep} propose deep end-to-end networks for lesion classification on digital mammography pictures. They also explore three different modes to develop deep CNN networks for whole mammogram diagnosis~\cite{zhu2017deep}. They apply their algorithm on the INbreast dataset, and the experimental results show the robustness of their networks~\cite{zhu2017deep}.

So far, mammography has been analyzed by much previous work with deep learning algorithms~\cite{litjens2017survey}. But there are little previous work that analyzes breast MRI, US, or digital breast tomosynthesis. In the near future, these other modalities will probably receive much attention~\cite{litjens2017survey}.

Because of most of the large digital datasets can not be used for free, many research efforts apply their algorithm on stale and small databases, which results in precarious AUC~\cite{sun2017enhancing}.  Much previous work has tackled this issue by employing semi-supervised learning~\cite{sun2017enhancing}, weakly supervised learning~\cite{paeng2017unified}, and transfer learning~\cite{kooi2017discriminating}.

\subsection{Drug discovery}
\textbf{Target identification and validation}
For a disease, determining its target is a challenging and time-consuming task.
Lamb et al.~\cite{lamb2006connectivity} create the first set of reference sets of gene expression profiles derived from cultured human cells treated with biologically active small molecules,
and pattern matching software to mine these data.
Madhukar et al.~\cite{madhukar2017new} develop a platform that integrates multiple data types into a Bayesian machine learning framework to predict the goals and mechanisms of small molecules.
They use publicly available BANDIT  data set  to achieve approximately 90\% accuracy on more than 2,000 different small molecules-significantly better than any other published target recognition platform.

\textbf{Retrosynthetic analysis}
Retrosynthetic analysis has a large search space, which is the biggest challenge of drug discovery.
Law et al.~\cite{law2009route} design a retrosynthetic analysis tool utilizing automated retrosynthetic rule generation.
However, the algorithm does not have good efficiency and effectiveness.
Inspired by alphgo, Segler et al.~\cite{segler2018planning} use Monte Carlo tree search (MCTS) and symbolic artificial intelligence to discover retrosynthetic routes.
The Waller team~\cite{segler2018planning} integrate concepts such as deep neural networks and reinforcement learning into a common architecture, and propose an algorithmic framework using three different neural networks together with MCTS(3N-MCTS).
Deep neural networks are used to predict which molecules will participate in the reaction. Monte Carlo search tree is used to predict the likelihood of a reaction.
Compared with traditional rule-based retrosynthesis analysis, this work borrows a lot of ideas from deep neural network and reinforcement learning, which is an important improvement to the traditional methods. 
The 3N-MCTS method is applied to the Reaxys database. Chemical reactions recorded before 2015 are used as training data, and chemical reactions recorded after 2015 are used as test data.
Compared with two traditional approaches: Heuristic BFS and Neural BFS, 3N-MCTS solves 87.12\% of the test set's problems while Neural BFS solves 45.6\% and Heuristic BFS solves 84.24\%~\cite{law2009route}~\cite{segler2018planning}.


\begin{table}[htbp]
  \caption{Summary of state-of-the-art work for drug discovery}
  \begin{center}
  \scalebox{0.78}[0.78]{
  \begin{tabular}{|p{0.45in}|p{0.9in}|p{1in}|p{0.5in}|p{0.2in}|}
  \hline
  \textbf{Reference} & \textbf{Techinique} & \textbf{Application} & \textbf{DB} & \textbf{Year} \\ \hline
  \cite{madhukar2017new} & Bayesian machine learning framework & target identification and validation & public data & 2017 \\ \hline
  \cite{law2009route} & automated retrosynthetic rule generation & retrosynthetic analysis & MOS & 2009 \\ \hline
  \cite{segler2018planning} & MCTS, DNN & retrosynthetic analysis &  Reaxys chemistry database & 2018 \\ \hline
  \cite{ramsundar2015massively} & Multitask Networks & drug screening & 259 datasets & 2015 \\ \hline
  \cite{wallach2015atomnet} & DCNN & bioactivity prediction & ChEMBL, DUDE & 2015 \\ \hline
  \end{tabular}
  }
  \label{tabdrugdiscoveryalg}
  \end{center}
  \begin{tablenotes}
    \small
    \item \textit{\textbf{Abbreviation:}} MOS = Accelrys and the Beilstein Crossfire reaction database; MCTS = Monte Carlo tree search; DNN = deep neural networks; DCNN = Deep Convolutional Neural Network;.
  \end{tablenotes}
\end{table}

\textbf{Drug screening}
In drug screening, virtual screening has been a hot topic because of its high time overhead of high-throughput screening and high cost of consumption.
The researchers at Google and Stanford~\cite{ramsundar2015massively} are working to develop virtual screening techniques using deep learning to replace or enhance traditional high-throughput screening processes and increase the speed and success rate of screening.
By applying deep learning, researchers can share information across numerous experiments across multiple targets. They  obtain data from 259 public datasets and divide the data into four groups, using a 5-fold cross-validation in group PCBA to achieve an AUC value of 0.873.
The 5-fold cross-validation AUC scores in the grouped MUV and Tox21 reach 0.841 and 0.818, respectively.
In drug screening, it is important to predict the ability of binding between molecules.
Wallach et al.~\cite{wallach2015atomnet} use structure-based deep-convolutional neural network - AtomNet to predict bioactivity.
They evaluate the accuracy of the model on the famous Directory of Useful Decoys Enhanced (DUDE) benchmark platform. AtomNet reaches or exceeds 0.9 AUC on 59 targets.
Two previous research efforts~\cite{ramsundar2015massively}~\cite{wallach2015atomnet} use state-of-the-art deep learning techniques to effectively reduce the cost of drug screening and improve accuracy.

In one word,  each step in drug discovery has relatively mature algorithms and platforms~\cite{costa2014big}.
We summarize the related algorithms in Table~\ref{tabdrugdiscoveryalg}.

\section{Are there any benchmarks for measuring algorithms, systems for big medical data?}
  ~\label{sec:bencmarks}

Benchmark is the foundation of systems and algorithms design, optimization, and evaluation. Many algorithms and systems have been widely used in medical domains. To explore processing efficiency and tackling potential bottlenecks, domain-specific benchmarks are essential for the developments and optimizations of that domain. However, in terms of big medical data domain, the diversity of disease taxonomies, the complexity of clinician research tasks, and the heterogeneity of medical data pose great challenges in constructing a comprehensive and fair benchmark.

An ideal medical benchmark should cover a broad spectrum of big medical data processing. To cover the data diversity, a benchmark should include electronic health records, laboratorial data, and QS data. To assure the algorithm diversity, a benchmark should consist of  not only traditional machine learning but also deep learning algorithms.

At present, there is no comprehensive big medical data benchmark suite. The previous benchmarking efforts only cover limited perspectives of big medical data. Table~\ref{tab611} compares different medical benchmarks from the perspectives of application domain, data type, data size, algorithm, system, metric and publishing year.

\begin{table*}[htbp]
\caption{Overview of medical data benchmarks.}
\begin{center}
\scalebox{0.78}[0.78]{
\renewcommand{\arraystretch}{1.5}
\begin{tabular}{|p{0.5in}|p{1in}|p{0.9in}|p{1.5in}|p{1.5in}|p{0.6in}|p{0.8in}|c|}
\hline
\textbf{Reference} & \textbf{Application Domain} & \textbf{Data Type} & \textbf{Data size} & \textbf{Algorithm} & \textbf{System} & \textbf{Metric} & \textbf{Year} \\ \hline
~\cite{liu2018benchmark} & Fall detection & sensor data & 9, 379 files  & the artificial neutral network, k nearest neighbor, support vector machine, and kernel Fisher discriminant & none & Accuracy, Specificity, Sensitivity & 2018\\ \hline
~\cite{rajpurkar2017mura} & abnormality detection in musculoskeletal radiographs & image data & 40,561 & 169-layer DenseNet baseline model & none & AUC, Specificity, Sensitivity & 2017\\ \hline
~\cite{zhan2016bigdatabench} & Gene & unstructed text data & gene sequence: 20MB-7GB, gene assembly: 100MB-13GB  & offline analysis & Work Queue, MPI & system and architecture metrics & 2016\\ \hline
~\cite{allen2016crowdsourced} & Alzheimer's Disease (AD) prediction &  Clinical data; Genotype data; Magnetic resonance image data & 767 training samples for question one; 176 CN samples for question two; 628 training samples for question three & AD prediction algorithms & none & Balanced accuracy, AUC & 2016 \\ \hline
~\cite{christov2010benchmark} & blood transfusion process & none & none & none & none & none & 2010\\ \hline
~\cite{shamir2008iicbu} &  biological image analysis & Image (TIFF format) & 9 datasets, 4,073 images & WND-CHARM multi-purpose image classification & none & Accuracy & 2008\\ \hline
~\cite{darling2003design} & nucleotide or protein sequences & Nucleotide, Peptide & 209,775,348 loci, 263,957,884,539 bases, from 209,775,348 reported sequences & parallel basic local alignment search & MPI & execution time & 2003\\ \hline
~\cite{sand} & genome assembly & Text & 4 datasets, 41,861,131 Reads & candidate filtering and alignment & Work Queue & none & -\\ \hline
\end{tabular}
}
\label{tab611}
\end{center}
\end{table*}

Liu et al.~\cite{liu2018benchmark} propose a benchmark database for fall detection. This database~\cite{liu2018benchmark} collects data from 50 males and females ranging from 21 to 60 years of age, 1.55 to 1.90 m in height, and 40 to 85 kg in weight. They use four baseline algorithms (ANN, KNN, SVM, and kernel Fisher discriminant) to evaluate the reliability of the database compared to the previous ones.

MURA~\cite{rajpurkar2017mura} is a benchmark database of musculoskeletal radiographs containing 40,561 multi-view radiographic images collected from 12,173 patients, with a total of  14,863 studies covering seven study types---elbow, finger, forearm, hand, humerus, shoulder, and wrist. Each study is labelled as normal or abnormal manually.

ChestX-ray8~\cite{wang2017chestx} is a hospital-scale chest X-ray database and benchmarks on weakly-supervised classification and localization of common Thorax diseases. It comprises 108,948 frontal-view X-ray images of 32,717 unique patients with the text-mined  eight  disease  image labels, from the associated radiological reports using natural language processing~\cite{wang2017chestx}.

AD DREAM Challenge~\cite{allen2016crowdsourced} is a benchmark suite aiming to evaluating the state-of-the-art algorithms in predicting AD, based on high dimensional, publicly available genetic and structural imaging data. The training data consist of individuals participating in the Alzheimer's Disease Neuroimaging Initiative (ADNI)~\cite{ad-challenge}.

Christov et al.~\cite{christov2010benchmark} present a medical benchmark based on a blood transfusion process.
The benchmark~\cite{christov2010benchmark} consists of a blood transfusion process definition, a set of properties or requirements, and a set of bindings between the blood transfusion properties and process definition.

IICBU 2008~\cite{shamir2008iicbu} is a benchmark suite for biological image analysis. It~\cite{shamir2008iicbu} provides a biological image datasets and a set of practical real-life imaging problems in biology, including the examples of organelles, cells and tissues. They can be used to evaluate different biological image analysis methods.

BigDataBench~\cite{wang2014bigdatabench,zhan2016bigdatabench} provides a big data and AI benchmark covering search engine, e-commerce, social network, multimedia, and bioinformatics domains. As for bioinformatics, it provides two workloads---SAND and BLAST. Among them, SAND~\cite{sand} is a set of modules for genome assembly. It performs distributed computing and supports easily deployments on large-scale clusters, clouds or grids. Totally, it consists of two steps including candidate filtering and alignment. As for datasets, it provides genome sequence data with three data scales---small, medium and large.
BLAST~\cite{darling2003design} is a parallel basic local alignment searching workload. It is used to compare nucleotide or protein sequences with the database and find the similarities.

\section{What is the performance gap of  state-of-practise and state-of-the-art systems for handling big medical data currently or in future?}
   ~\label{sec:gaps}

Many state-of-the-art and state-of-practise systems have been widely used in big medical data. However, the characteristics of medical data pose great challenges to both data storage and processing.

Multiple medical data processing systems are proposed to handle large-scale medical data. For medical imaging data analysis, PAIS (Pathology Analytical Imaging Standards)~\cite{foran2011imageminer} provides a data model to manage image data, using a spatial DBMS based architecture. Hadoop-GIS~\cite{wang2011hadoop} adopts a MapReduce-based solution to support complex queries for analytical pathology imaging.
For genome data analysis, GATK~\cite{mckenna2010genome} is a MapReduce-based  genome analysis toolkit for analyzing next-generation DNA sequencing data. IMG~\cite{markowitz2011img} is an integrated microbial genomes database and comparative analysis system. For medical text data, Neamatullah et al.~\cite{neamatullah2008automated} provide a system to process text-based patient medical records.

However, the sources and types of medical data are usually multifarious and integrated. For example, the electronic health record (EHR) dataset, which is used in a recent research effort~\cite{rajkomar2018scalable}, has thousands of feature dimensions and contains patient demographics, provider orders, diagnoses, procedures, medications, laboratory values, vital signs, and flowsheet data~\cite{rajkomar2018scalable}, covering images, text, structured, or un-structured data types and sources. These dimensions are not processed and learned individually, and conversely, they are combined to detect and diagnose diseases cooperatively. Under this circumstance, the storage and processing systems are required to integrate different data sources and types. Existing systems targeting a specific data type like medical images can not process other data types.
To the best of our knowledge, there exists no such a system that can support multi-source and heterogeneous data storage and processing in the big medical domain. Previous work~\cite{rajkomar2018scalable} develops a new data structure---FHIR standard to handle their heterogeneous data. However, considering the diversity of disease taxonomies and the variety of medical data, developing a new data structure or a new system for one or more data types or data sources only provide a case-by-case solution. Hence, there is an urgent need for a new comprehensive system that satisfies the processing requirements of big medical data.

Nowadays, deep learning systems, e.g. TensorFlow~\cite{abadi2016tensorflow}, Caffe~\cite{jia2014caffe}, are used to handle medical data. However, since these systems are general-purpose deep learning processing framework without specific optimizations for big medical data, they have inefficiencies considering complex clinician research tasks. As illustrated above, EHR dataset~\cite{rajkomar2018scalable} has thousands of correlative feature dimensions covering images, text, structured, or un-structured data types and sources, so the researchers not only need to conduct time-consuming data conversion so as to suit for the data input requirements of the above systems, but also need to correlate those  data manually to collect all the data for each patient.

\section{Are we ready for working together?}
   ~\label{sec_collaboration}

The tendency to utilize computing technologies in medicine science is indispensable in the future for two reasons: on one hand, a large amount of high quality data has been accumulated in the medicine fields with the fast-paced medical development. On the other hand, machine learning techniques  have performed a good job in routine tasks, and the researchers of both fields have  already made progress in many aspects of medicine, i.e., \emph{automatic} disease diagnosis.

However, we notice that the practical and wide application of machine learning techniques to clinical therapy is limited by several significant problems: first, the development of data-driven AI in the directions of clinical medicine has been restricted by the low degree of clinical practice digitization and the difficulty of sharing data. Second, clinical practices cover exceedingly complex scenarios while current machine learning techniques merely have a good performance in the scenarios where the conditions are described explicitly. 
Thus, we believe that several prerequisites should be met with  for the sake  of an effective and efficient multidisciplinary cooperation. First, the  digitized degree of clinical practice and the sharing of clinical data should be improved because these two issues limit the computer experts to deepen their understanding on clinical medicine. Second, the unified criteria of the case utilizing AI or other machine learning techniques in clinical tasks should be set up to help experts of both fields to validate and share the knowledge among  each other. Benchmarks will play an unique role in setting up those unified criteria of the case utilizing AI in clinical tasks.
Third, the multidisciplinary education among computing  and medicine sciences should be promoted to provide sufficient talent with comprehensive abilities.

\section{conclusion}
~\label{sec_conclusion}

In this paper, we---a group of life scientists,  clinicians, computing scientists and engineers perform a comprehensive survey on big medical data, which is heterogeneous, high-dimensional, embodying a large mixture of signals and errors, and  significantly different from other domain data. We investigate the prioritized tasks in clinician practices and research: quantified self (QS), disease classification, disease diagnosis, and drug discovery.  We found  publicly availably data sets that can be utilized for those tasks are not limited to its scale, but also its single data source. The previous work demonstrates the potential of incorporating machine learning techniques into clinician practices. However, its high accuracy is achieved on the static data.  In reality, the clinician practitioners work in an open environments, so we need set up the realistic benchmarks that can mimic the way that the clinician practitioners handle the data for different medical data purposes.
The sources and types of medical data are usually multifarious and integrated. These dimensions are not processed and learned individually, and conversely, they are combined to detect and diagnose diseases cooperatively.
 To the best of our knowledge, there exists no such a system which can support multi-source and heterogeneous
data storage and processing in the big medical domain. Also, we discuss several prerequisites for the sake  of an effective and efficient  cooperation among life scientists,  clinicians, computing scientists and engineers.

\section*{Acknowledgment}

This work is supported by the National Key Research and Development Plan of China (Grant No. 2016YFB1000600, 2016YFB1000605, and 2016YFB1000601).


\begin{IEEEbiography}[{\includegraphics[width=1in,height=1.25in,clip,keepaspectratio]{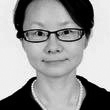}}]{ Zhifei Zhang} received the M.S. degree in physiology and the Ph.D. degree in etiology from Capital medical University Beijing, China, in 2003 and 2011, respectively. She is currently a researcher in  Beijing Key Laboratory of respirology, Capital medical university, China. Her research focuses on medical big data and circulatory physiology.
\end{IEEEbiography}

\begin{IEEEbiography}[{\includegraphics[width=1in,height=1.25in,clip,keepaspectratio]{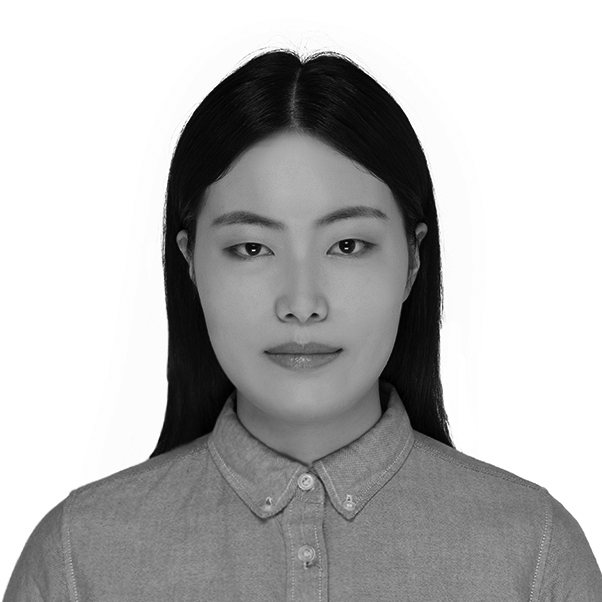}}]{Wanling Gao} is an assistant professor in computer science at the Institute of Computing Technology, Chinese Academy of Sciences and University of Chinese Academy of Sciences from 2019. Her research interests focus on Big Data and AI benchmarking and systems. She received her B.S. degree in 2012 from Huazhong University of Science and Technology, and Ph.D. degree in computer science and engineering from University of Chinese Academy of Sciences in 2019, respectively.
\end{IEEEbiography}

\begin{IEEEbiography}[{\includegraphics[width=1in,height=1.25in,clip,keepaspectratio]{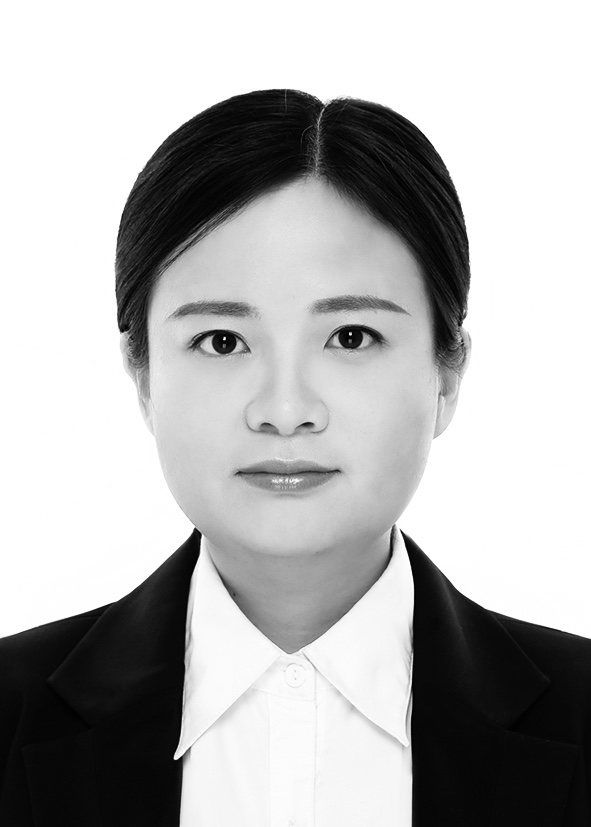}}]{Fan Zhang} received the B.S. degree and M.S. degree in computer science from Xi'an Jiaotong University, Xi'an, China, in 2013 and 2016, respectively. She is currently an engineer of State Key Laboratory of Computer Architecture, Institute of Computing Technology, Chinese Academy of Sciences, Beijing. Her research focuses on data mining.
\end{IEEEbiography}

\begin{IEEEbiography}[{\includegraphics[width=1in,height=1.25in,clip,keepaspectratio]{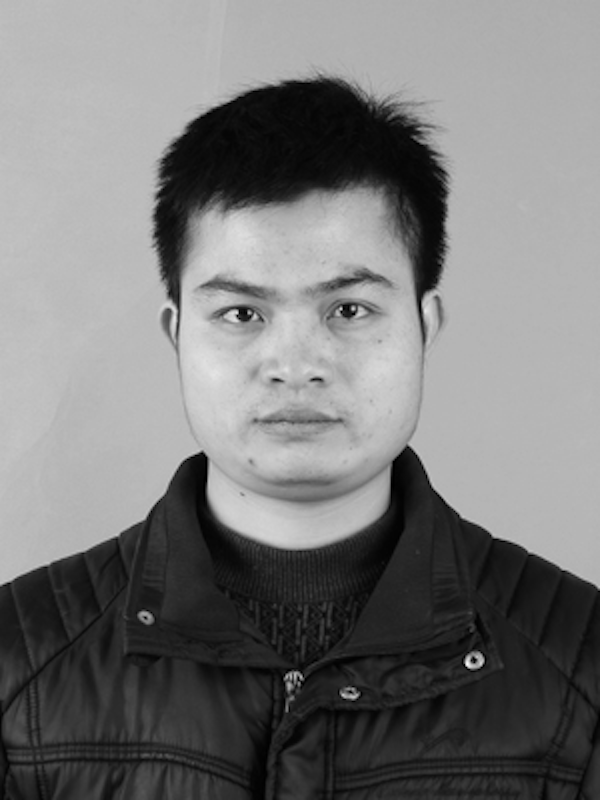}}]{Yunyou Huang} received the B.S. degree and M.S. degree from the Guangxi Normal University, Guilin, China, in 2012 and 2015, respectively. He is currently pursuing Ph.D. degree in Institute of Computing Technology (ICT), Chinese Academy of Sciences (CAS) and University of Chinese Academy of Sciences. His research interests include data analytics in smart grid and machine learning.
\end{IEEEbiography}

\begin{IEEEbiography}[{\includegraphics[width=1in,height=1.25in,clip,keepaspectratio]{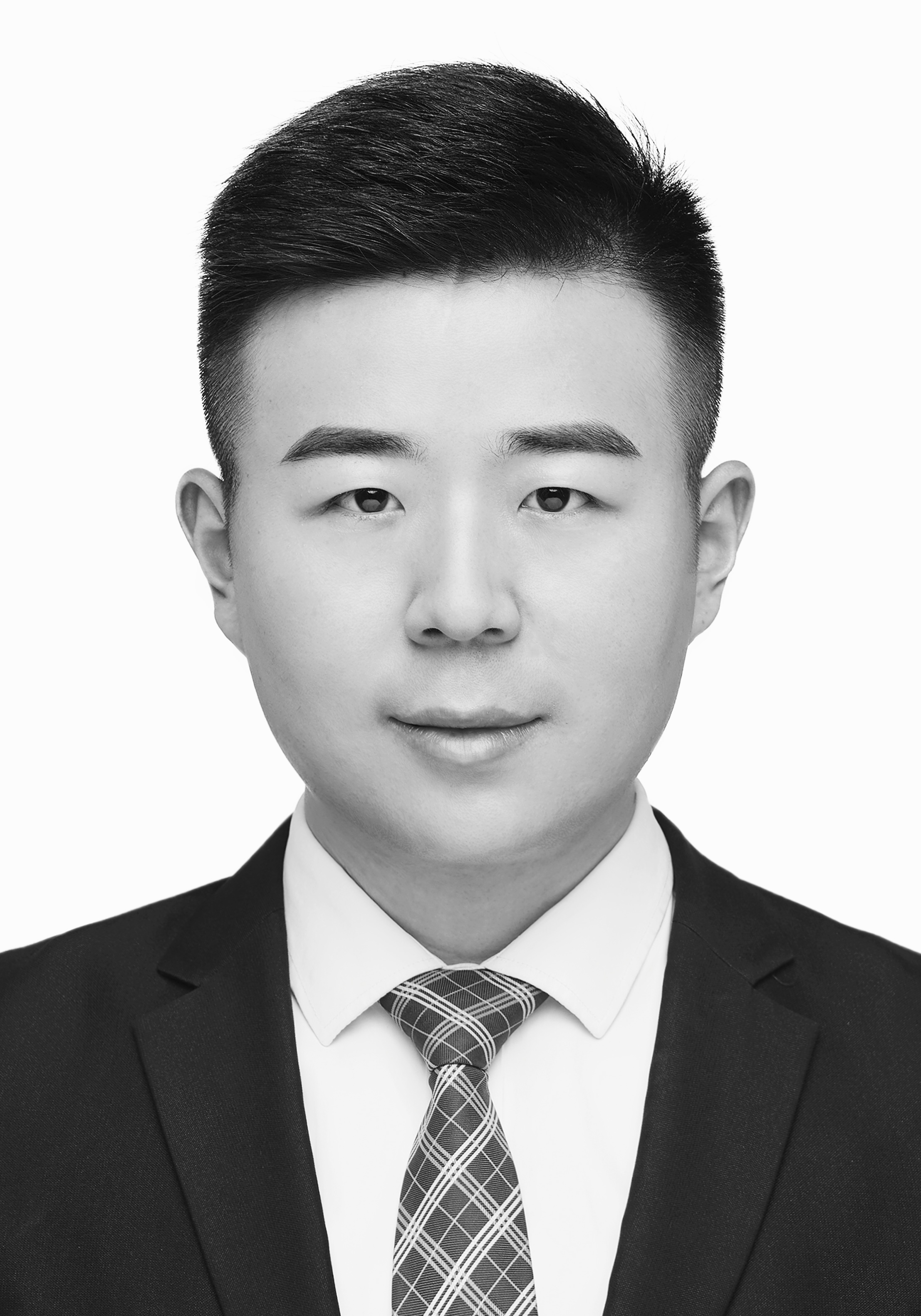}}]{Shaopeng Dai} is currently pursuing the Ph.D. degree in computer science at the Institute of Computing Technology, Chinese Academy of Sciences and University of Chinese Academy of Sciences. His research interests focus on medical data analysis, big data benchmark and deep learning. He received the B.S. degree in computer science and technology from Beijing University of Posts and Telecommunications, Beijing, China, in 2013.
\end{IEEEbiography}

\begin{IEEEbiography}[{\includegraphics[width=1in,height=1.25in,clip,keepaspectratio]{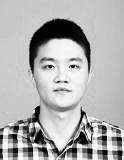}}]{Fanda Fan} received the B.S. degree from the School of Software Engineering, Sichuan University, Chengdu, China, in 2017. He is currently pursuing the Master's degree with the State Key Laboratory of Computer Architecture, Institute of Computing Technology, Chinese Academy of Sciences, Beijing.
His research interests include data mining, and machine learning.
\end{IEEEbiography}

\begin{IEEEbiography}[{\includegraphics[width=1in,height=1.25in,clip,keepaspectratio]{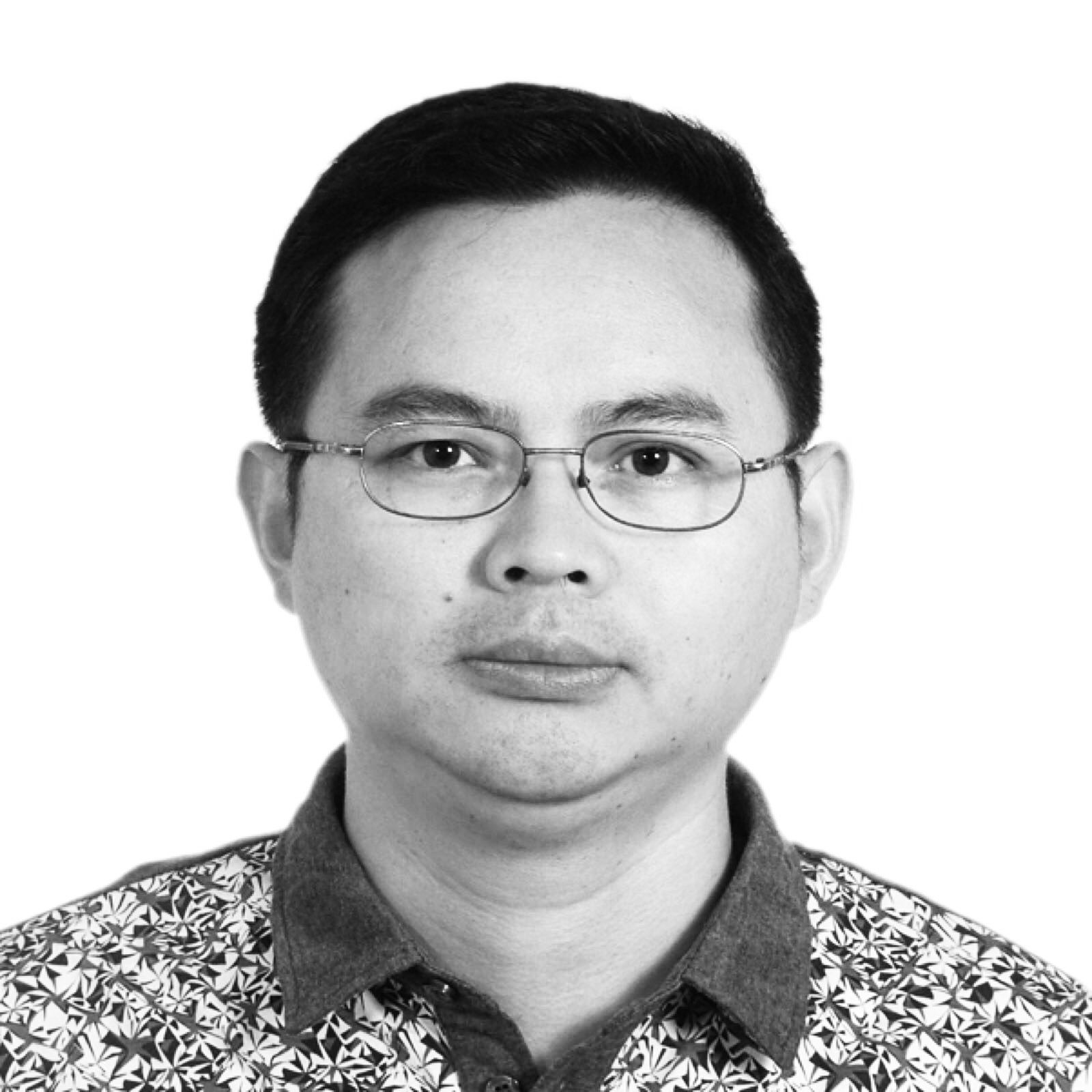}}]{Jianfeng Zhan} received the B.S. degree in civil engineering and the M.S. degree in solid mechanics from Southwest Jiaotong University, Chengdu, China, in 1996 and 1999, respectively, and the Ph.D. degree in computer software and theory from the Institute of Software, Chinese Academy of Sciences, and University of Chinese Academy of Sciences in 2002. Since 2012, he has been a Full Professor with the State Key Laboratory of Computer Architecture, Institute of Computing Technology, Chinese Academy of Sciences, and University of Chinese Academy of Sciences. He has authored over 150 articles. He has been working on systems, architecture, and algorithms for more than 20 years. He has served as executive committee chair of BenchCouncil---a multi-disciplinary international open benchmarking council (http://www.benchcouncil.org), and TPDS associate editor.
\end{IEEEbiography}

\begin{IEEEbiography}[{\includegraphics[width=1in,height=1.25in,clip,keepaspectratio]{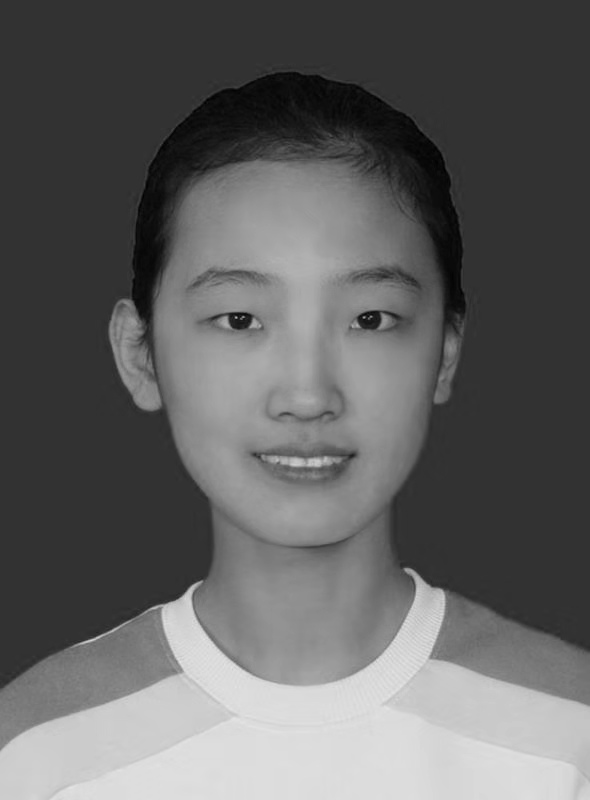}}]{Mengjia Du} received the B.S. degree from the School of Software, Nankai University, Tianjin, China, in 2013. She is currently pursuing the academic master's degree with the State Key Laboratory of Computer Architecture, Institute of Computing Technology, Chinese Academy of Sciences, Beijing.
Her research interests include medical image recognition and clinically assisted diagnosis.
\end{IEEEbiography}

\begin{IEEEbiography}[{\includegraphics[width=1in,height=1.25in,clip,keepaspectratio]{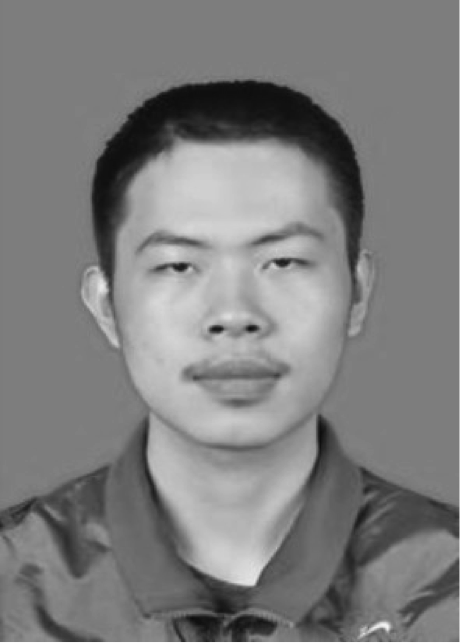}}]{ Siling Yin} is a guest student in Institute of Computing Technology, Chinese Academy of Sciences, Beijing. His research interests include machine learning, big data. He received the B.S. degree from the College of Information Science and Technology, Beijing University of Chemical Technology, Beijing, China, in 2018.
\end{IEEEbiography}

\begin{IEEEbiography}[{\includegraphics[width=1in,height=1.25in,clip,keepaspectratio]{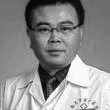}}]{ LONGXIN XIONG} received the M.S. degree in surgery from NanChang University Jiangxi, China, in 2005. He is an associate professor of Nanchang First Hospital, Jiangxi, China. He has been engaged in clinical work of general surgery for more than twenty years and is skilled in the treatments of benign and malignant liver tumors, gastrointestinal benign and malignant tumors, benign or malignant thyroid tumors, benign and malignant breast tumors.
\end{IEEEbiography}

\begin{IEEEbiography}[{\includegraphics[width=1in,height=1.25in,clip,keepaspectratio]{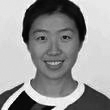}}]{ JUAN DU} received the M.S. degree in clinical medicine from Capital medical University Beijing, China, in 2009. She is an attending doctor in Neonatal Center Beijing Children's Hospital Affiliated to Capital Medical University. Her research focuses on Prematurity Neonatal nutrition Neonatal respiratory disease.
\end{IEEEbiography}

\begin{IEEEbiography}[{\includegraphics[width=1in,height=1.25in,clip,keepaspectratio]{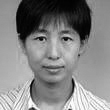}}]{ YUMEI CHENG}  received the M.S. degree in clinical medicine from Capital Medical University Beijing, China, in 2003. She is an attending doctor in  Beijing Obstetrics and Gynecology Hospital, Affiliated to Capital Medical University. Her research focuses on Gynecology.
\end{IEEEbiography}

\begin{IEEEbiography}[{\includegraphics[width=1in,height=1.25in,clip,keepaspectratio]{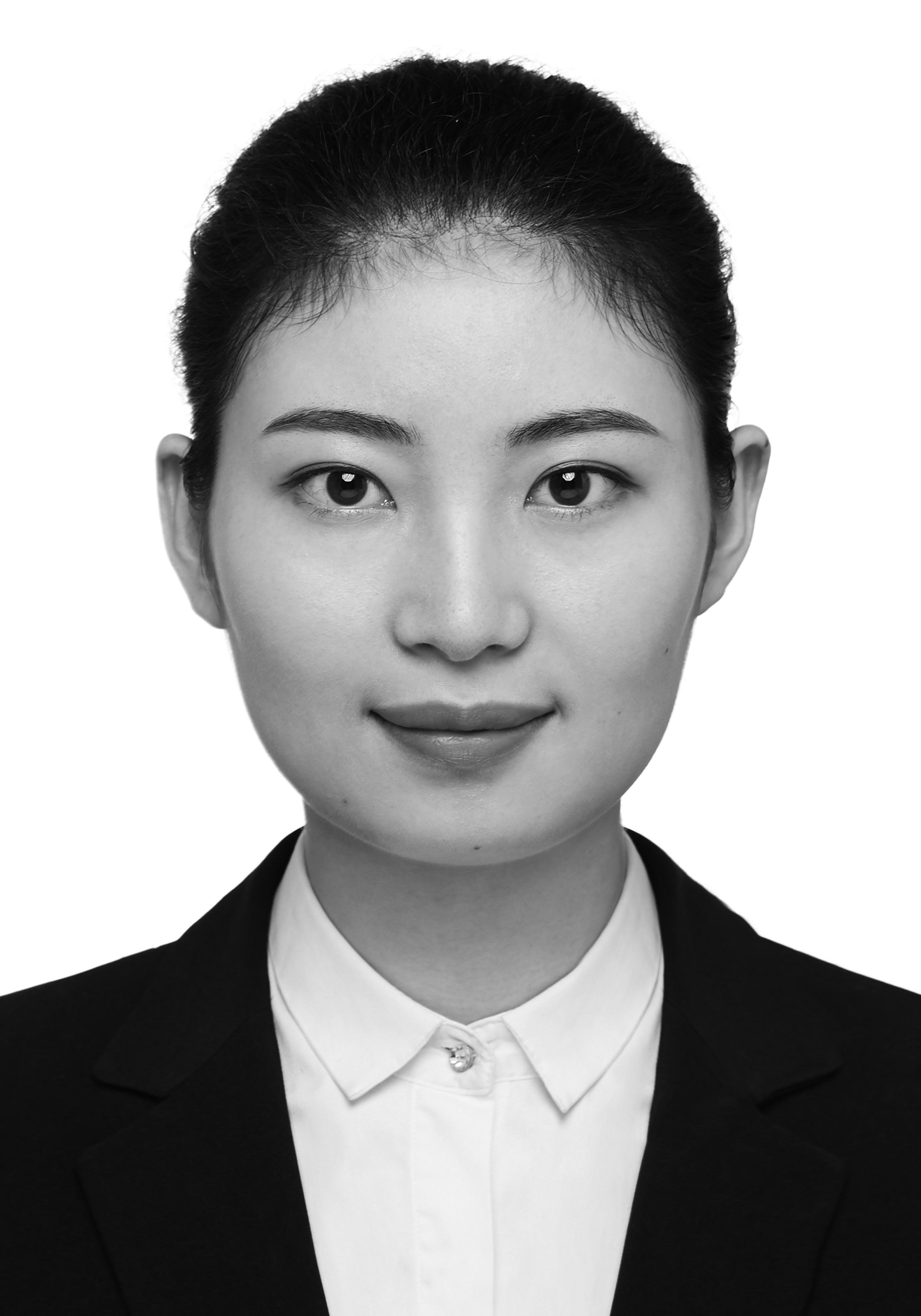}}]{ Xiexuan Zhou}received the B.S. degree from the School of Computer Science and Technology, Beijing University of Posts and Telecommunication, Beijing, China, in 2014. She is currently pursuing the Ph.D. degree with the State Key Laboratory of Computer Architecture, Institute of Computing Technology, Chinese Academy of Sciences, Beijing.
Her research interests include big data of mass spectrum, deep learning and transfer learning.
\end{IEEEbiography}

\begin{IEEEbiography}[{\includegraphics[width=1in,height=1.25in,clip,keepaspectratio]{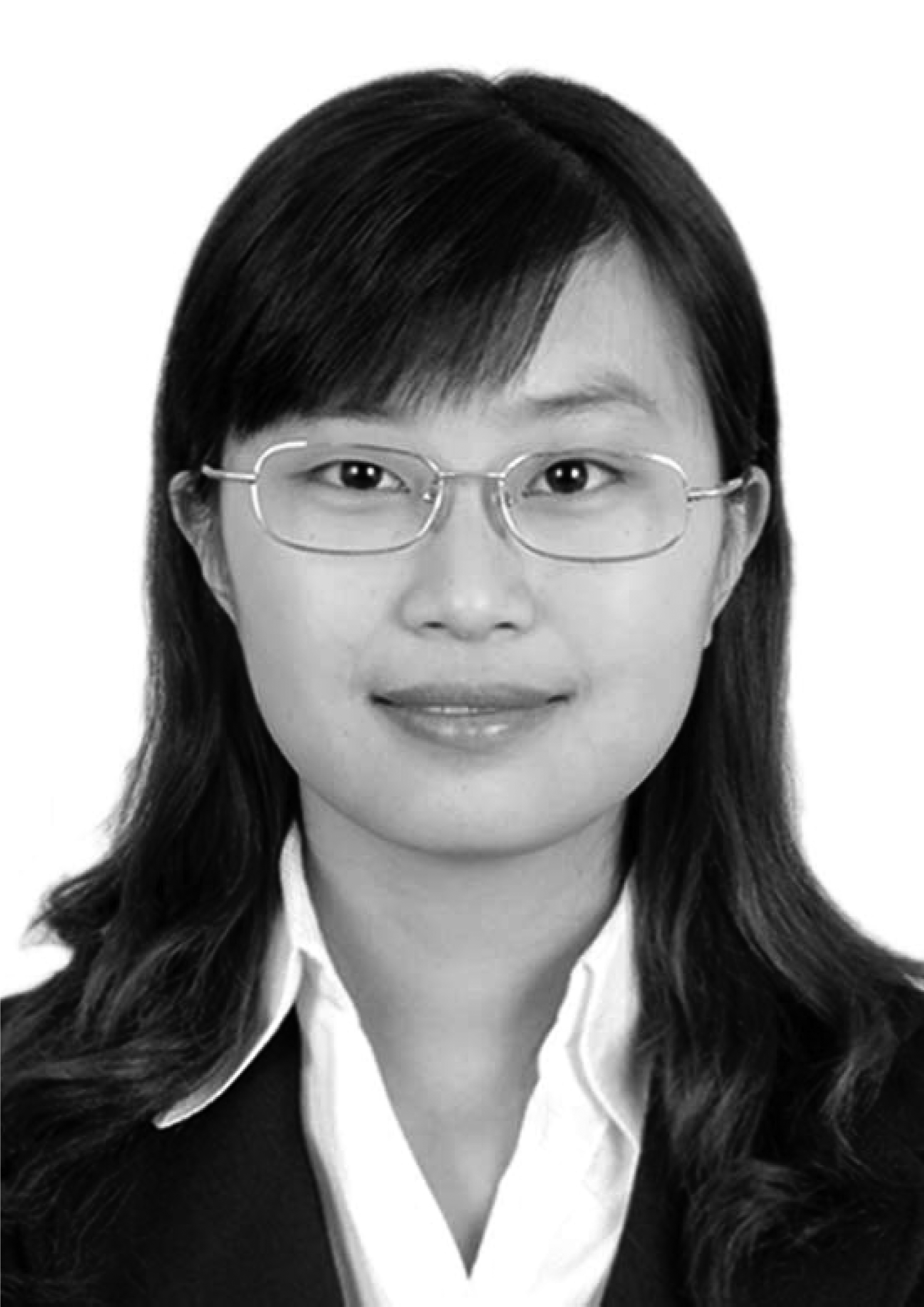}}]{Rui Ren} received the BS degree in computer science from the SiChuan University, China, in 2009, the MS degree in computer architecture
from Chinese Academy of Sciences, Beijing, China, in 2012. She is currently a Ph.D student and engineer in the Institute of Computing Technology, Chinese Academy of Sciences. Her research
interests include big data, distributed systems, performance analysis of cluster and cloud systems.
\end{IEEEbiography}

\begin{IEEEbiography}[{\includegraphics[width=1in,height=1.25in,clip,keepaspectratio]{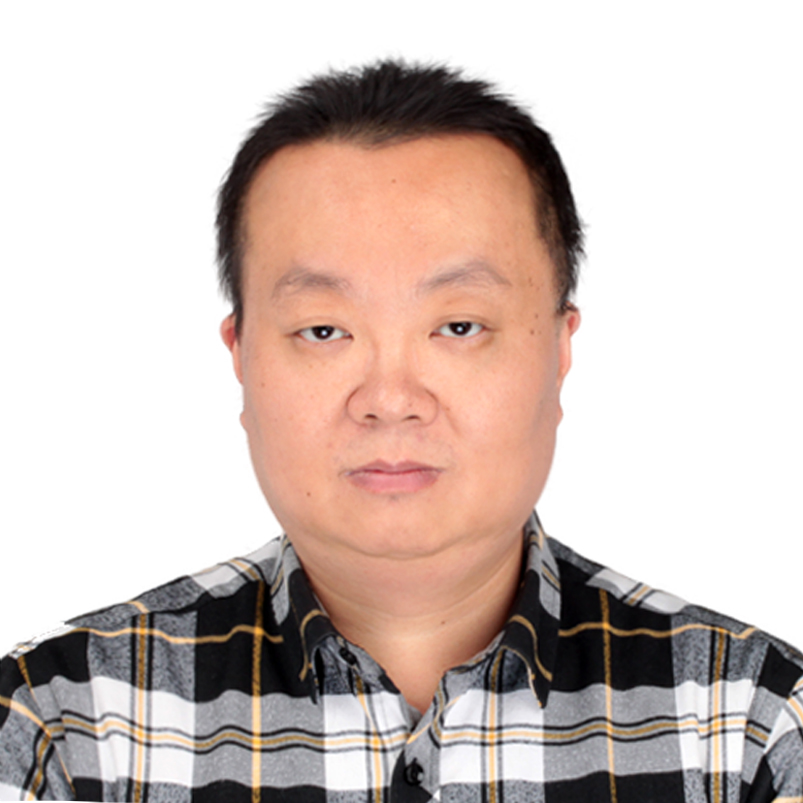}}]{ Lei Wang} received the B.S. degree in applied mathematics from Beijing University of Technology, Beijing, China, in 1999, and the M.S. degree in computer engineering and the Ph.D. degree in computer software and theory from University of Chinese Academy of Sciences, Beijing, China, in 2006 and 2016, respectively.  He is currently a senior engineer with Institute of Computing Technology, Chinese Academy of Sciences. His current research interests include benchmarking and resource management of cloud systems.
\end{IEEEbiography}

\begin{IEEEbiography}[{\includegraphics[width=1in,height=1.25in,clip,keepaspectratio]{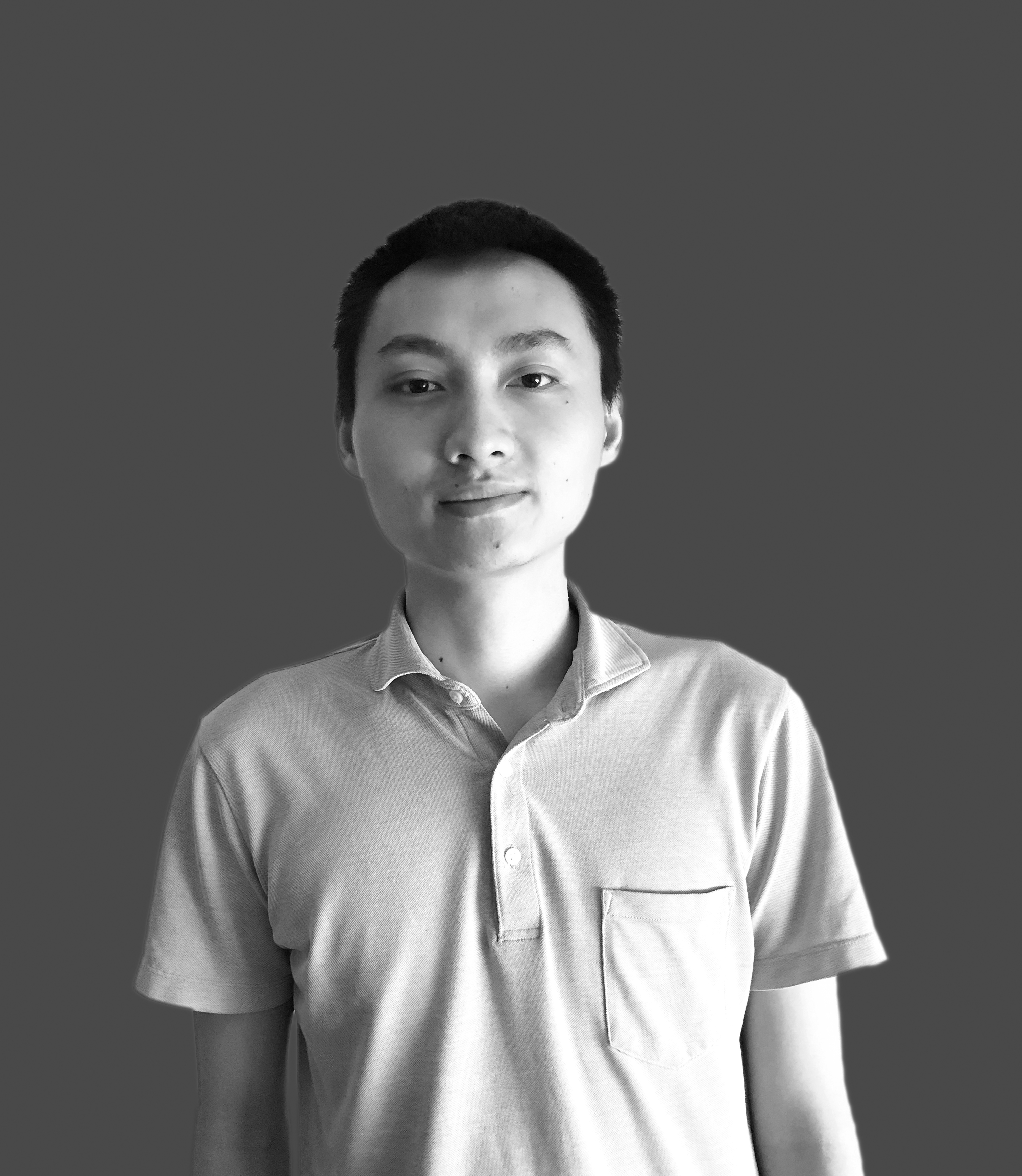}}]{Hainan Ye} received the BS degree in computer science from Beihang University in 2015. He is a manager at Beijing Academy of Frontier Sciences and Technology.
\end{IEEEbiography}

\EOD

\end{document}